\documentclass[epj,nopacs]{svjour}
\usepackage{graphicx}
\usepackage{color}
\usepackage[numbers,sort&compress]{natbib}
\usepackage{amsmath}
\usepackage{hyperref}

\allowdisplaybreaks[4]

\begin{document}
%
\title{Search for CP violation effects in the $h\to \tau\tau$ decay with $e^+e^-$ colliders}
\author{Xin Chen\inst{1,}\inst{2,}\inst{3,}\thanks{\email{xin.chen@cern.ch}}
\and Yongcheng Wu\inst{1,}\inst{4,}\thanks{\email{ycwu@physics.carleton.ca}} 
}                     
%
%
\institute{Department of Physics, Tsinghua University, Beijing 100084, China \and Collaborative Innovation Center of Quantum Matter, Beijing 100084, China \and Center for High Energy Physics, Peking University, Beijing 100084, China \and Ottawa-Carleton Institute for Physics, Carleton University, 1125 Colonel By Drive, Ottawa, ON K1S 5B6, Canada}
\date{Received: date / Revised version: date}
%
\abstract{
A new method is proposed to reconstruct the neutrinos in the $e^+e^-\to Zh$ process followed by the $h\to\tau\tau$ decay. With the help of a refined Higgs momentum reconstruction from the recoiling system and the impact parameters, high precision in the determination of the momentum of neutrinos can be achieved. 
The prospect of measuring the Higgs CP mixing angle with the $h\to\tau\tau$ decay at $e^+e^-$ colliders is studied with the new method. The analysis is based on a detailed detector simulation of the signal and backgrounds. The fully reconstructed neutrinos and also other visible products from the tau decay are used to build matrix element (ME)-based CP observables. With 5 $ab^{-1}$ of data at $E_{\text{CM}}=250$ GeV, a precision of $2.9^\circ$ can be achieved for the CP mixing angle with three main one-prong decay modes of the taus. The precision is found to be about 35\% better than the other methods.
%
} 
\maketitle
\section{Introduction}
\label{sec:intro}

To explain the matter-antimatter asymmetry in the current observed Universe, C and CP symmetry violation is one of the three necessary conditions listed by Sakharov~\cite{Sakharov}. In the electroweak baryogenesis models, the CP violation phase
from the CKM matrix is too small to accommodate the observed baryon-antibaryon imbalance. Hence the Standard Model has been extended to introduce extra new CP violating sources, such as in the 2HDM, Left-Right Symmetric and SUSY models. 

On the other hand, the recently discovered Higgs~\cite{ATLAS-Higgs,CMS-Higgs,higgs_mass} has opened a new window toward the new physics. The precision measurement of the Higgs properties will be one of the most important targets of the next generation collider including high-luminosity upgrade of LHC~\cite{CERN-LHCC-2015-020} and $e^+e^-$ colliders~\cite{ILC-TDR,Behnke:2013lya,CEPC-CDR,vanderKolk:2017urn}. Besides the determination of the signal strengths of the Higgs for different channels~\cite{Peskin:2012we,Klute:2013cx,Han:2013kya}, the large data that will be accumulated in future colliders would allow us to measure the CP structure of some certain interaction with extraordinary precision. 

Using 8 TeV LHC data, ATLAS and CMS have already measured the CP property of the Higgs coupling with the $Z$ boson using the four-lepton decay channel ($h\to ZZ\to(\ell\ell)(\ell'\ell')$). A pure CP odd assumption is excluded at 99\% confidence level~\cite{Aad:2013xqa,Chatrchyan:2012jja,Chatrchyan:2013mxa}.
On the other hand, a CP violating Higgs sector can also be searched in the fermionic decays of Higgs, using an effective Lagrangian of the following form:
\begin{equation}
\mathcal{L} =  -\frac{y_\tau}{\sqrt{2}} ( \cos\phi\bar{\tau}\tau + i \sin\phi\bar{\tau}\gamma_5\tau) h ,
\label{eq:eq1}
\end{equation}
where $\phi$ is the mixing angle between the CP even and odd terms. The advantage of this decay is that, unlike the bosonic decays of Higgs where the pseudoscalar interaction may arise from a dimension-6 operator such as $h Z_{\mu\nu}\tilde{Z}^{\mu\nu}$, the pseudoscalar term in Eq.~\ref{eq:eq1} is dimension-4 and the coefficient can be large, so that a large CP violating effect is possible when the Higgs is a CP mixture. Although difficult to reconstruct, the tau decays are complex enough to yield information of the tau spin correlation~\cite{Hagiwara:1989fn,Bullock:1992yt}. Hence $h\to\tau^+\tau^-$ is a golden channel for the measurement of the CP nature of the Higgs which has led to an enormous amount of studies on the prospects of the measurement of the CP mixing angle($\phi$) at LHC using either the ggF/ZH or the VBF production mode~\cite{Berge:2008wi,Berge:2008dr,Berge:2011ij,Harnik:2013aja,Dolan:2014upa,Berge:2015nua, Askew:2015mda,Hagiwara:2016zqz,Han:2016bvf,He:1993fd} as well as at $e^+e^-$ colliders~\cite{Bower:2002zx,Desch:2003mw,Berge:2013jra}. 

The tau decay will always produce at least one neutrino which can hardly be reconstructed at hadron colliders.\footnote{In Ref.~\cite{Hagiwara:2016zqz}, the authors have provided a method to reconstruct the neutrinos in the $\tau\to\pi\nu_\tau$ channel at LHC by using the PDF of $\vec{p}_T^{\text{miss}}$.} Hence, the proposed methods at LHC utilize either only the visible products (such as the four pions in the $\rho\rho$ mode) or the impact parameter of the tau~\cite{Berge:2015nua}, all of these can be reconstructed without knowing the full information of the neutrino. However, the resolution of the hadron collider will significantly affect the reconstruction of the CP sensitive observable~\cite{Han:2016bvf}. Furthermore, besides the clean background, the lepton collider can also provide the possibility to fully reconstruct the neutrinos. Thus the lepton collider will be a better facility for the study of the CP properties of the Higgs.

In this paper, we propose a new method that can be used to reconstruct the momentum of the neutrinos from tau decay at future $e^+e^-$ machine. In this method we utilize all the information including mass peaks of all involving particles as well as the impact parameters of each tau decay which will be measured with higher precision at lepton collider than at hadron collider. We find that using the new proposed method we can precisely determine the momentum of the missing neutrinos. With these fully reconstructed neutrinos as well as the visible decay products we construct a CP sensitive observable based on event by event matrix element to test the CP structure of the coupling between Higgs and $\tau$ leptons. After performing a detailed Monte Carlo (MC) simulation of the signal and corresponding backgrounds at $e^+e^-$ collider, we have checked capability of the matrix element-based observable and also compared with the usually used CP observable defined as the angle between some certain planes.

The rest of the paper is organized as follows. In Sect.~\ref{sec:mainmethod} we present in detail the method that we used to reconstruct the neutrino. We first refine the Higgs momentum from recoiling $Z$ boson, and then determine the momentum of neutrinos by a global fitting utilizing impact parameter and on-shell conditions. A detailed MC simulation study of this method is given in Sect.~\ref{sec:MC}. We use a matrix element-based CP observable to give the prospect of the measurement of the CP mixing angle. A comparison with previously used method is also given in Sect.~\ref{sec:MC}. We summarize our discussion in Sect.~\ref{conclusion}.

\section{The reconstruction of neutrinos at \texorpdfstring{$e^+e^-$}{e+e-} collider}
\label{sec:mainmethod}

The leading Higgs production channel at $e^+e^-$ collider is 
\begin{eqnarray}
e^+ e^- \to Z h.
\end{eqnarray}
In order to study the CP property of the $h\tau\tau$ coupling, after the Higgs decays into $\tau$ pair (including the CP violation effect as stated in Eq.~(\ref{eq:eq1})), and the $Z$ boson decays either leptonically ($Z\to\ell\ell$, $\ell=e/\mu$) or hadronically ($Z$  $\to jj$), the following one-prong decay channels for $\tau$-leptons will be used:
\begin{eqnarray}
\tau^\mp &\to& \nu_\tau\ \ell^\mp\ \nu_\ell, \nonumber \\
\tau^\mp &\to& \nu_\tau\ \pi^\mp, \nonumber \\
\tau^\mp &\to& \nu_\tau\ \rho^\mp\to\nu_\tau\ \pi^\mp\ \pi^0. 
\end{eqnarray}
For clarity, before introducing the reconstructed method, we will first introduce details of the simulation and basic reconstruction for signal and background processes.
The charged particle in each process leaves a track inside the detector which is used to construct the impact parameters. The $\tau$ mass will be used in the reconstruction of the neutrino, as well as the $\rho$ mass in the $\rho$ decay channel. Further, the mass of the Higgs will also be used to provide more information. Note that one can construct the Higgs peak from either the momentum of $\tau$ pair or the recoil mass constructed by the momentum of the $Z$ boson, and how well the Higgs momentum can be reconstructed will significantly affect the further determination of the neutrino momentum. Thus the Higgs momentum reconstruction from the recoil system will also be investigated.

\subsection{Signal and background simulation and reconstruction}
\label{simu_reco}

The signal and background events are generated with MadGraph5~\cite{MG5} and passed to Pythia8~\cite{Pythia8} for the resonance decay (CP mixing Higgs, tau and rho mesons) and parton shower. The spin correlation between two taus is retained during the whole procedure. The signal process as demonstrated above is $e^+e^-\to Zh$ ($\sigma=212$ fb), at a center of mass energy $E_{\text{CM}}=250$ GeV, with $h\to\tau\tau$ and $Z\to \ell\ell$ or $Z\to jj$(2 jets). The dominant backgrounds for this signal are following:
\begin{itemize}
\item $e^+e^-\to ZZ,\ Z\to\tau\tau,\ Z\to \ell\ell(jj)$ 
\item $e^+e^-\to Zh,\ Z\to\tau\tau,\ h\to bb$
\item $e^+e^-\to Zh,\ Z\to\tau\tau,\ h\to l\nu l\nu (l=e/\mu/\tau)$
\end{itemize}
The events are afterwards passed through the DELPHES~\cite{Delphes} simulation using the ILD card based on the TDR of ILC~\cite{Behnke:2013lya}, in which the tracking efficiency for charged tracks is $99\%$ with fiducial pseudorapidity range up to $|\eta|=2.4$; the identification efficiency for electrons and muons is $95\%$, the calorimeter towers are simulated with a particle flow algorithm with a coverage up to $|\eta|=3.0$. The track momentum is smeared with a resolution of $\sqrt{0.01^2 + (10^{-4}p_T)^2}$ ($p_T$ in GeV) in a magnetic field of 3.5 T, and a resolution of 0.001 in $\eta$ and $\phi$ (azimuthal angle) for its direction. The calorimeter energy is smeared with a resolution of $\sqrt{A^2E^2 + B^2E}$, where the coefficients for the constant and stochastic terms are $A=1.0\%$ ($1.5\%$) and $B=15\%$ ($50\%$), respectively, for the EM (hadronic) calorimeter.
To reject non-prompt leptons from jet fragmentation and heavy flavor meson decay, a lepton isolation cut of $\sum p_T^{\text{PF}}/p_T^{\ell}<0.7$ is applied, where $\sum p_T^{\text{PF}}$ is the sum over particle flow objects (tracks and calorimeter clusters) around the lepton with $p_T>0.5$ GeV and $\Delta R=\sqrt{\Delta\eta^2+\Delta\phi^2}<0.4$ with respect to the lepton. This efficiency of this cut ranges from $91\%$ in the low $p_T^{\ell}$ (5 GeV$<p_T^{\ell}<10$ GeV) to $99\%$ in the high $p_T^{\ell}$ ($>30$ GeV) region, and mostly affects the soft leptons from tau decays.

To have the best impact parameter resolutions which will be used to reconstruct the neutrinos, a minimum $p_T$ of 5 GeV is applied on the leptons and charged pions from taus. The resolution of the impact parameter in the transverse plane (in the $z$-axis) is set to $\sigma_d=5$ $\mu$m ($\sigma_z=10$ $\mu$m) as in Eq.~\ref{eq:eq6}, which can be achieved in the next generation $e^+e^-$ colliders~\cite{Behnke:2013lya,CEPC-CDR}.

The hadronic taus ($\tau_h$) are clustered by the Anti-$k_{t}$ jet algorithm~\cite{antikt} with a cone parameter of 0.4. A simple flat $\tau$-tagging efficiency of $60\%$ ($0.5\%$) is applied on the real (fake) taus, taking into account the current $\tau$-tagging performance at the LHC experiments~\cite{tau_id2, tau_id1}. With this fake tau rate, the $W^+W^-\to 4j$ background with two jets faking two taus can be neglected. When the $Z$ boson decays hadronically, the particle flow objects associated with the selected electrons, muons and hadronic taus are firstly removed from the event, and then a coneless exclusive ($k_t$) algorithm~\cite{kt} is applied on the remaining objects to cluster up to two jets, which are assumed to originate from $Z\to jj$. The four-momentum of $Z$ can be best reconstructed in this way, which further helps to calculate the Higgs mass recoiling against $Z$ with the given CM energy. The efficiency of finding two jets with $p_T>5$ GeV and $|\eta|<3$ is about $95\%$.

The net efficiencies after the above object selections in different di-tau decay modes and the requirement that leptons from the $h$/$Z$ decay should have opposite charge are listed in Tab.~\ref{tab:tab2}. To measure the CP mixing angle, it is essential to identify the different tau decay modes efficiently. The development of tau substructure algorithms in the ATLAS and CMS experiments has recently made this possible based on the particle flow method~\cite{tausub1,tausub2}. The neutral pion energy can also be measured with a precision of $15\%$ with the substructure. In this work, we assume that different tau decay channels can be efficiently identified in the future $e^+e^-$ colliders, with no crosstalk for simplicity, and a precision of $10\%$ on the neutral pion energy measurement can be achieved.
\begin{table}[tb]
\centering
\begin{tabular}{c|ccccc} \hline
  & $\ell+\pi$ & $\ell+\rho$ & $\pi+\pi$ & $\pi+\rho$ & $\rho+\rho$ \\ \hline
$Z\to ee/\mu\mu$ & $31.4\%$ & $27.2\%$ & $19.2\%$ & $18.5\%$ & $15.7\%$ \\ \hline
$Z\to jj$ & $34.8\%$ & $30.8\%$ & $24.5\%$ & $21.3\%$ & $18.9\%$ \\ \hline
\end{tabular}
\caption{The combined efficiencies of selecting the leptons, taus and jets for the different $Z$ and ditau decay modes (neutrinos are omitted in the notation).}
\label{tab:tab2}
\end{table}


\subsection{Refined Higgs four-momentum reconstruction}
\label{sec:higgs_4p}
\begin{figure*}[tb]
\centering
\includegraphics[width=0.32\textwidth]{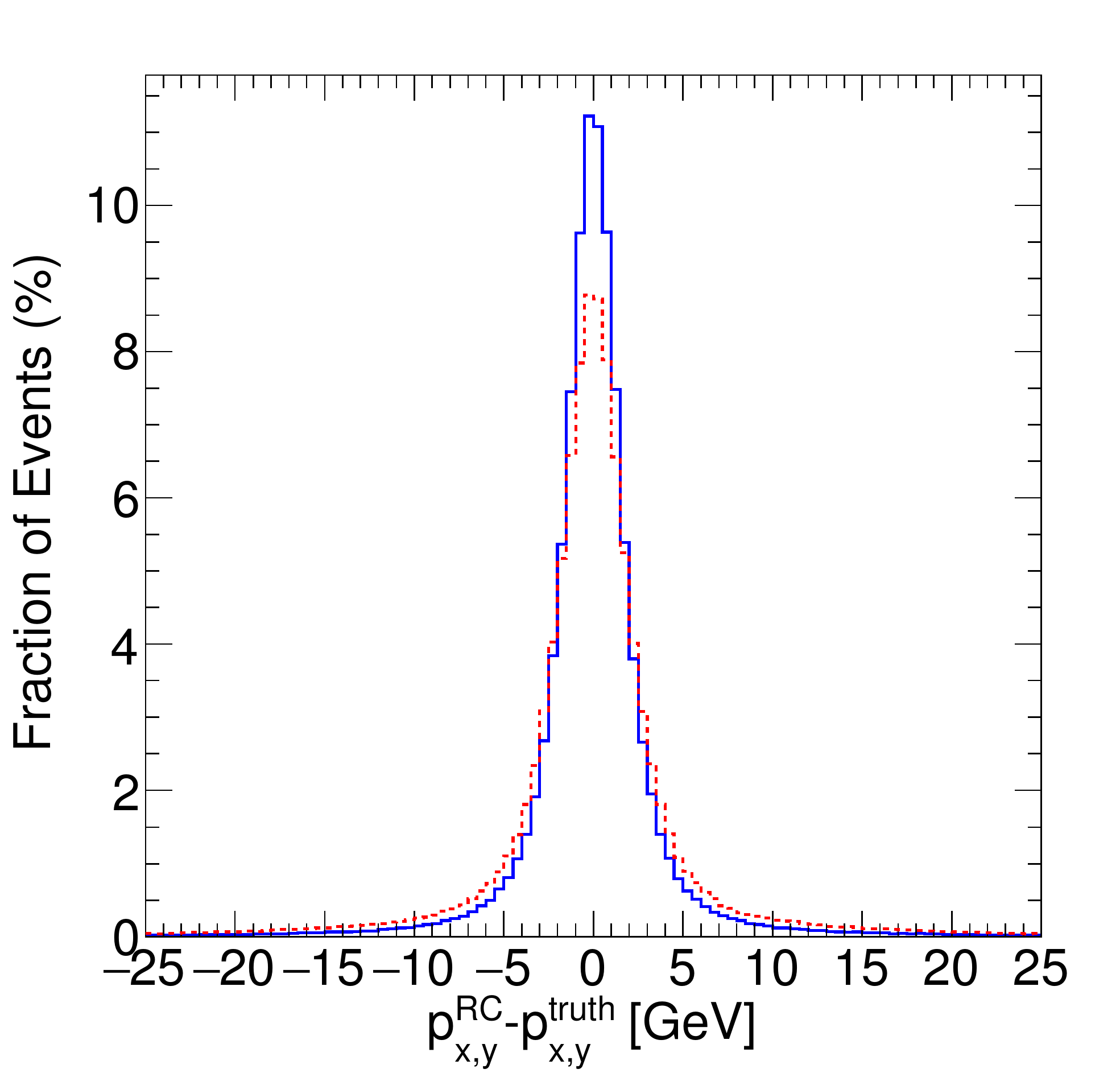}\hspace{2mm}
\put(-40, 100){\textbf{(a)}}
\includegraphics[width=0.32\textwidth]{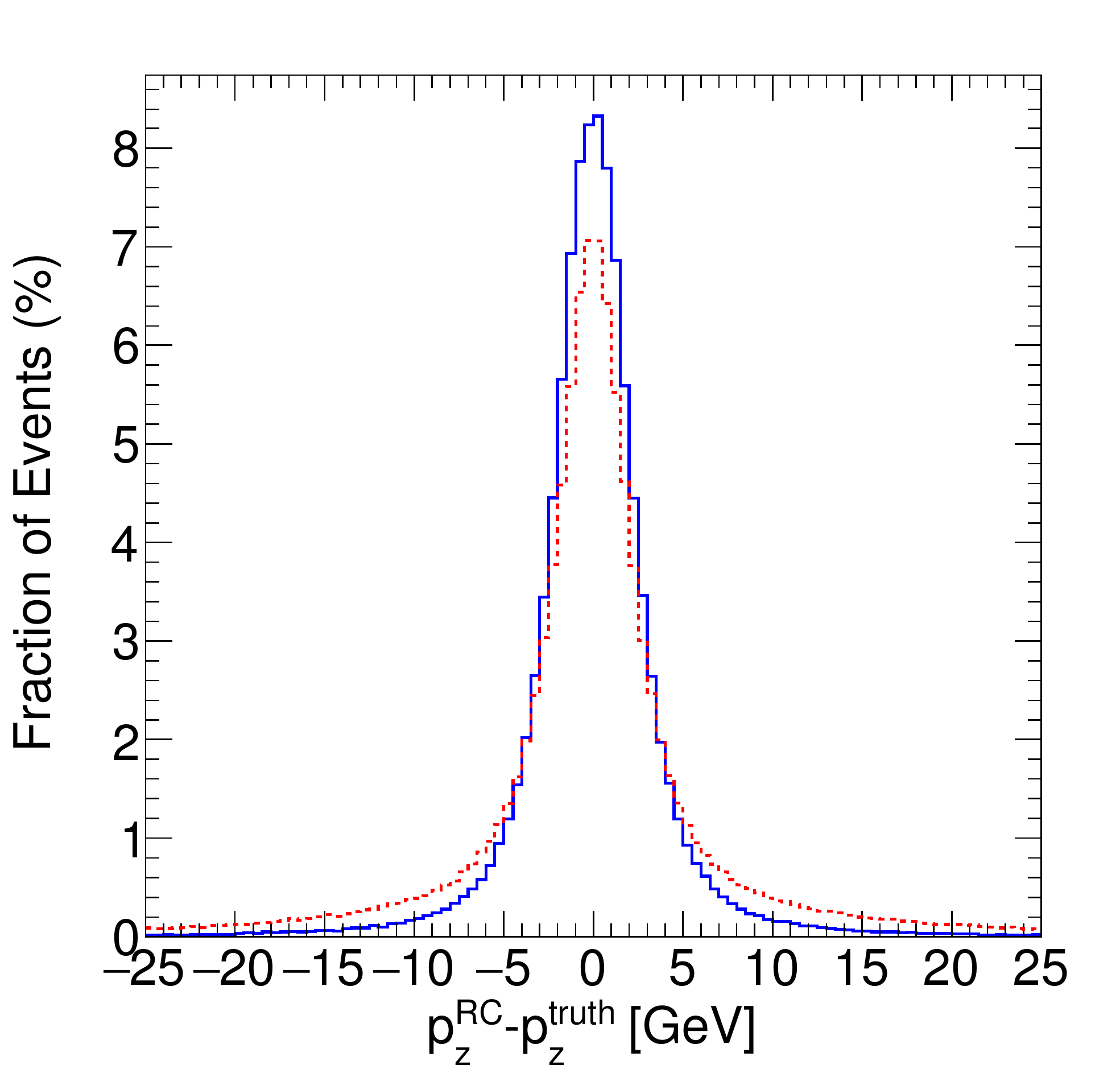}\hspace{2mm}
\put(-40, 100){\textbf{(b)}}
\includegraphics[width=0.32\textwidth]{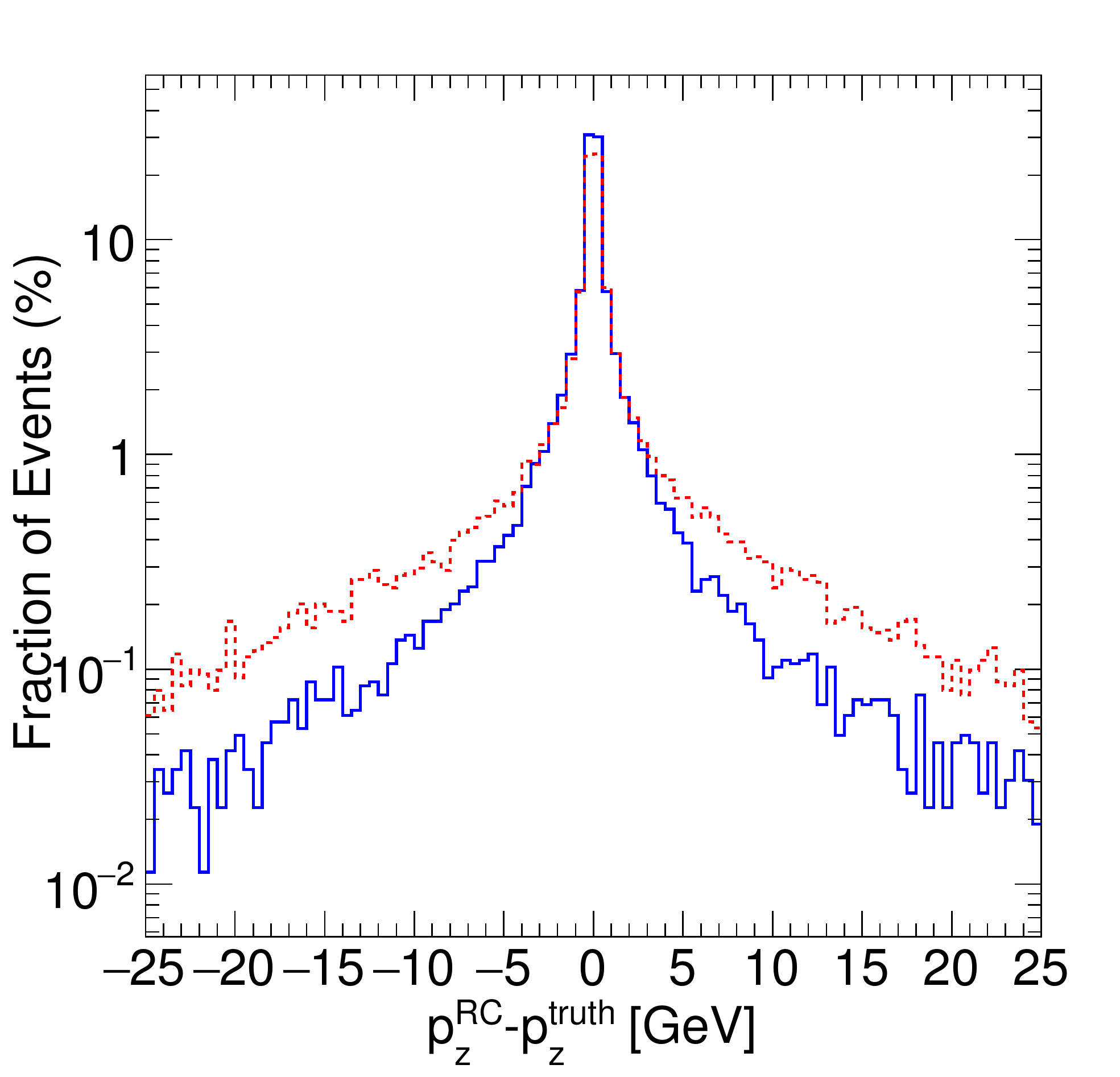}
\put(-40, 100){\textbf{(c)}}
\caption{\label{fig:recoil} The resolutions of the Higgs recoil momentum in $x$ and $y$ (a) and in $z$ (b) in the $Z\to jj$ channels, and in $z$ in the $Z\to \ell\ell$ channels (c). The red dashed (blue solid) histogram corresponds to the calculation based on the original uncorrected momenta of jets or leptons from $Z$ (improved momenta of jets from from $Z$ and the consideration of Eq.~\ref{eq:eq2}). }
\end{figure*}

In general, the reconstructed Higgs peak will always have a long tail as we ignored the initial state radiation (ISR) photons. 
To best reconstruct the Higgs four-momentum from the recoiling $Z$ boson, taking into account the known Higgs mass at 125 GeV, and assuming that the ISR photons are mostly collinear with the beam, a quantity $x$ can be solved which represents the fraction of momentum carried away by the ISR photon from the positron beam traveling in the positive $z$-direction,
\begin{equation}
x = \frac{E_{\text{CM}}^2-2E_{\text{CM}}E+m^2-m_h^2}{\pm E_{\text{CM}}^2 \mp E_{\text{CM}}E+E_{\text{CM}}p_z},
\label{eq:eq2}
\end{equation}
where $E$, $m$ and $p_z$ are the energy, mass and $z$-component of the momentum of the recoiling $Z$ boson. The value of $x$ can also be negative, which means that the ISR photon carries momentum from the electron beam along the negative $z$-axis.~\footnote{Two hard ISR photons are rare and not considered.} With the $x$ solved, the four-momentum of the $Zh$ system can be expressed as $p_{\text{tot}}$ = ($E$, $p_x$, $p_y$, $p_z$) = (250-125$|x|$, 0, 0, -125$x$) GeV, from which the Higgs recoil four-momentum can be calculated as $p_h^{\text{RC}}$ = $p_{\text{tot}}$ - $p_Z$. Because the tau mass is much smaller than the Higgs mass, the tau from Higgs decay is highly boosted and the neutrinos from tau decay are almost collinear with its visible decay products. With this assumption, the Higgs four-momentum can also be expressed as $p_h$ = $p_{\text{vis1}}/x_1$ + $p_{\text{vis2}}/x_2$, where $p_{\text{vis1},2}$ are the four-momenta of visible decay products of the two taus, and $x_{1,2}$ are the fractions of momentum carried by the visible products. With these quantities defined, a $\chi^2$ can be minimized per event to get the best momentum resolution:
\begin{eqnarray}
\begin{array}{ll}
  \chi^2 = & \sum_{i=0}^{3}\left(\frac{p_{h,i}-p_{h,i}^{\text{RC}}}{0.5}\right)^2 + \left(\frac{m_Z-91.2}{2.5}\right)^2\\
  & + \left(\frac{f_{j1}-1}{0.06}\right)^2 + \left(\frac{f_{j2}-1}{0.06}\right)^2,
\end{array}
\label{eq:eq3}
\end{eqnarray}
where $m_Z$ is the $Z$ boson mass constructed from the decay products of the $Z$ boson (two leptons or two jets), $f_{j1,2}$ are the jet energy correction factors on the two jets from $Z\to jj$ decay, and variables with energy dimensionality are in GeV. The jet energy resolution used here is about 6\% and this is a conservative choice and can be further improved in future electron colliders, such as ILC~\cite{ILC-TDR,Behnke:2013lya}, CEPC~\cite{CEPC-CDR} and CLIC~\cite{vanderKolk:2017urn} (4\%). By minimizing the $\chi^2$ in Eq.~\ref{eq:eq3} with $x_{1,2}$ and $f_{j1,2}$ freely floating, not only the two-fold ambiguity of $x$ in Eq.~\ref{eq:eq2} is resolved (the solution with a smaller $\chi^2$ is chosen), but also the jet resolution, and hence the resolutions of $p_h^{\text{RC}}$ are improved. The improvement is shown in Fig.~\ref{fig:recoil}. The denominator values in Eq.~\ref{eq:eq3} are chosen such that the best resolutions as in Fig.~\ref{fig:recoil} are achieved. In the $Z\to \ell\ell$ channels, the last two terms in Eq.~\ref{eq:eq3} are not present since the leptons' momenta are precisely measured when compared to the jets. Eq.~\ref{eq:eq3} fulfills two purposes in this case. One is to resolve the $x$ ambiguity, and the other is to resolve the two-fold ambiguity of the $e^+e^-e^+\tau_h$ and $\mu^+\mu^-\mu^+\tau_h$ final states. There are two ways to assign the two same-sign leptons to the $Z$ and $h$ decays. With current setup, the fraction of wrong assignments is negligible.

\subsection{Impact parameter}
\label{sec:imp}
Ideally, except for the lepton channels, using the missing momentum, the Higgs four-momentum and the tau mass, one already has enough constraints to reconstruct the two neutrinos. However, in practice, the reconstructed momentum of the neutrino still has large uncertainty. Thus the usage of the further information from the impact parameters \cite{imp1,imp2} will be absolutely helpful in the reconstruction of the neutrinos especially in lepton channel. In order to make use of the impact parameters, a $\chi_{\text{IP}}^2$ is reconstructed for each tau which will be minimized in further global fittings; this will be explained in the next section. 

\begin{figure*}[tb]
\centering
\includegraphics[width=0.55\textwidth]{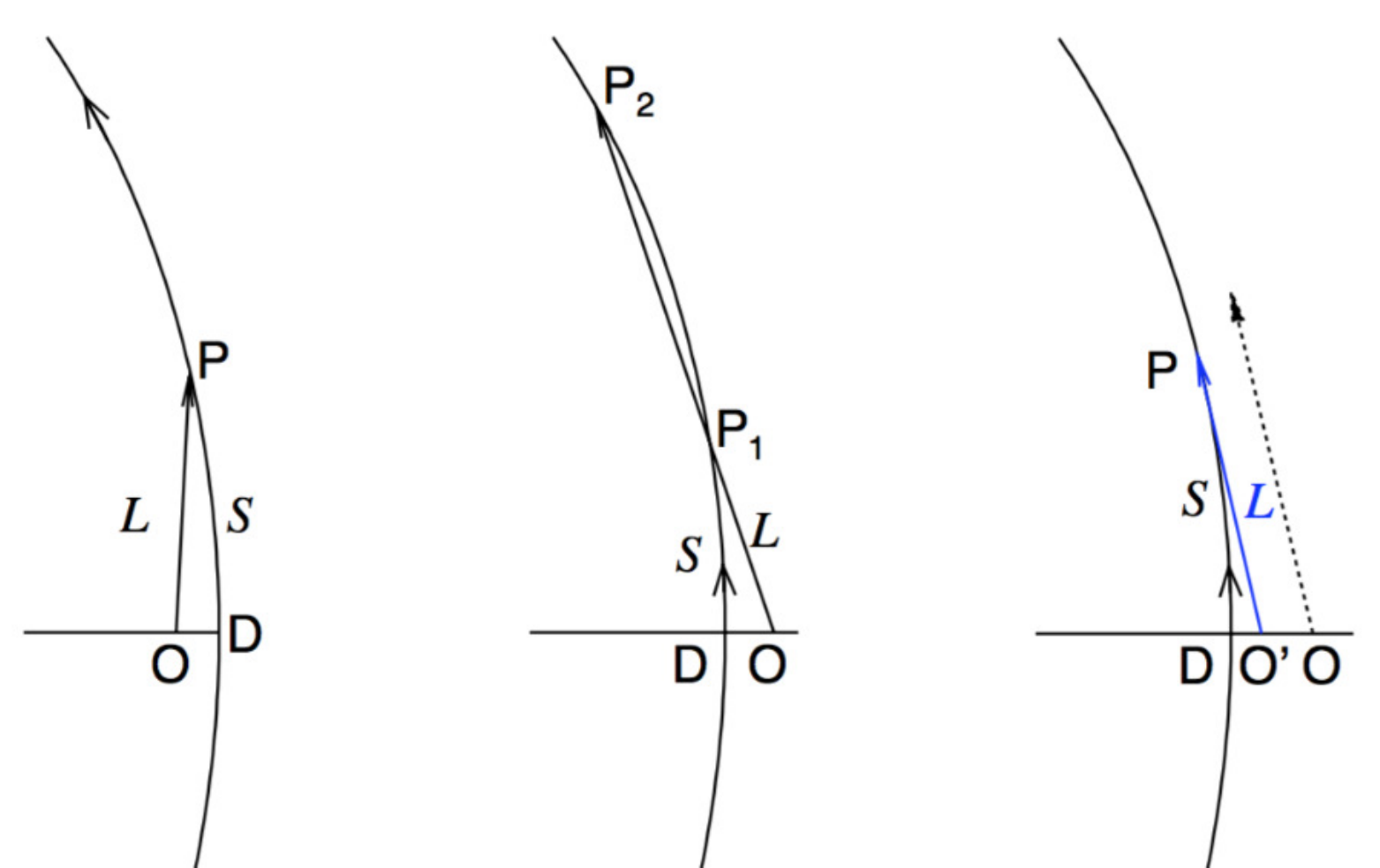}
\put(-190, 120){\textbf{(a)}}
\put(-110, 120){\textbf{(b)}}
\put(-20, 120){\textbf{(c)}}
\caption{\label{fig:imp} Plots demonstrating the tau flight vector, the track trajectory and the point of closest approach for three cases: (a) the collision point $O$ is inside the track curvature, (b) $O$ is outside the track curvature with two intersection points, (c) $O$ is outside the track curvature with no intersection. }
\end{figure*}

The method to calculate $\chi_{\text{IP}}^2$ for one tau is as follows: given the transverse impact parameter $d$ and the direction of tau momentum in the transverse plane, the intersection point between the tau flight direction and the track trajectory in the transverse plan can be found. The transverse impact parameter $d$ is defined in the transverse plane as the minimum distance from the interaction point to the charged track. As demonstrated in Fig.~\ref{fig:imp}, the tau is produced at the collision point $O$, flight length from $O$ to the point of decay $P$ in the transverse plane is $L$, the point of closest approach to $O$ with a backward extrapolation of the trajectory is $D$, and the extrapolated arc length in transverse plane is $S$. There is just one intersection point when $O$ is inside the circle of the trajectory (Fig.~\ref{fig:imp}(a)), but when $O$ is outside, the number of intersection points can be 2, 1 or 0 (Fig.~\ref{fig:imp}(b,c)). The distance between $O$ and $D$ is the impact parameter $d_0$. The impact parameter in the $z$-axis can be calculated from $P$: 

\begin{equation}
z_0 = L \sinh \eta_\tau - S \sinh \eta_{\text{track}}.
\label{eq:eq5}
\end{equation}

The impact parameters will get updated values ($d_0^{\text{fit}}$ and $z_0^{\text{fit}}$) when $\chi_{\text{IP}}^2$ is minimized. When just one intersection point exists, the relevant contribution reads

\begin{equation}
\chi^2 = \left(\frac{d_0^{\text{fit}}-d_0}{\sigma_d}\right)^2 + \left(\frac{z_0^{\text{fit}}-z_0}{\sigma_z}\right)^2.
\label{eq:eq6}
\end{equation}

When two intersection points exist, the $\chi^2$ for each possibility is calculated according to Eq.~\ref{eq:eq6} and the smaller one is taken. When no intersection point exists, as demonstrated by the dashed arrow in Fig.~\ref{fig:imp}(c), it is assumed to be due to the uncertainty in $d_0$ measurement, and the best-fit $d_0^{\text{fit}}$ will be around the distance between $O'$ and $D$ (denoted $d_0^{\text{C}}$). The $O'$ is determined by translating the dashed vector to become tangential to the trajectory (the blue vector), and $z_0^{\text{fit}}$ is calculated from the tangential point $P$. The relevant contribution then reads

\begin{equation}
\chi^2 = \left(\frac{d_0^{\text{fit}}+d_0-2d_0^{\text{C}}}{\sigma_d}\right)^2 + \left(\frac{z_0^{\text{fit}}-z_0}{\sigma_z}\right)^2 .
\label{eq:eq7}
\end{equation}

When the dashed vector in Fig.~\ref{fig:imp}(c) points away from the track curvature, no tangential point can be found, and $O'$ coincides with $D$. Although the tracks in Fig.~\ref{fig:imp} all travel anti-clockwise, but depending on the particle charge, they may also travel clockwise, i.e., from $P$ to $D$. Situations with tracks traveling clockwise can be obtained by taking a mirror image of the plots in Fig.~\ref{fig:imp}, with essentially no change to Eq.~\ref{eq:eq6} and \ref{eq:eq7}. The cases where the angle between the tau flight direction and the track trajectory is an obtuse angle are also considered in our method. However, since the tau is boosted and its decay products are highly collinear, this rarely happens. 

\subsection{Reconstruction of the neutrino momenta}
\label{neut_reco}

In order to best estimate the neutrino momenta, with the benefit of accurate particle momentum reconstruction and clean background from a $e^+e^-$ collider, the direction (in $\eta$, $\phi$) and the magnitude of each of the two neutrinos from tau decays are scanned globally for each event. The available information such as the tau mass, the Higgs four-momentum from $Z$ recoil, and the impact parameters of the charged tracks from tau decays, are all used to achieved this goal. With the four-momenta of all final state particles reconstructed, a matrix element-based method is used to fully extract the information of CP, which can result in an improved sensitivity with respect to the usually used method~\cite{Berge:2015nua}. 

The global $\chi^2$ of the likelihood that is minimized for every event is expressed as
\begin{eqnarray}
\begin{array}{ll}
  \chi^2 = & \sum_{i=0}^{3}\left(\frac{p_{h,i}-p_{h,i}^{\text{RC}}}{\sigma_{\text{RC}}}\right)^2 + \left(\frac{m_{\tau 1}-1.777}{\sigma_{\tau 1}}\right)^2\\
  & + \left(\frac{m_{\tau 2}-1.777}{\sigma_{\tau 2}}\right)^2 + \chi_{\text{IP}}^2,
\end{array}
\label{eq:eq4}
\end{eqnarray}
where $p_h = p_{\text{vis1}}+p_{\text{mis1}}+p_{\text{vis2}}+p_{\text{mis2}}$ is calculated from the four-momenta of visible and invisible decay products of the two taus (without any collinear assumptions), $p_{h}^{\text{RC}}$ is obtained by minimizing Eq.~\ref{eq:eq3} in the previous step, $m_{\tau 1,2}$ are the masses of the taus, and $\chi_{\text{IP}}^2$ is the term accounting for the contribution from the impact parameters of the tracks that can help the fit to find the correct neutrino directions as stated in Sect.~\ref{sec:imp}. For $Z\to \ell\ell$ ($jj$) channels, $\sigma_{RC}=0.5$ (4.0) GeV and $\sigma_\tau =0.1$ (0.2) GeV. The resolution parameters are set so as to achieve the best reconstructed neutrino momenta (magnitude and direction close to the true values). For the states with an intermediate $\rho$ meson, extra terms of $(m_\rho - 0.775)^2/\sigma_\rho^2 + (f_\rho - 1)^2/0.10^2$ are added to Eq.~\ref{eq:eq4}, where $f_\rho$ is the factor multiplied to the $\rho$ meson's energy for a better resolution. Correspondingly, $\sigma_\tau$ is doubled for this case, and $\sigma_\rho=0.15$ (0.30) for $Z\to \ell\ell$ ($jj$).

\begin{figure*}[tb]
\centering
\includegraphics[width=0.36\textwidth]{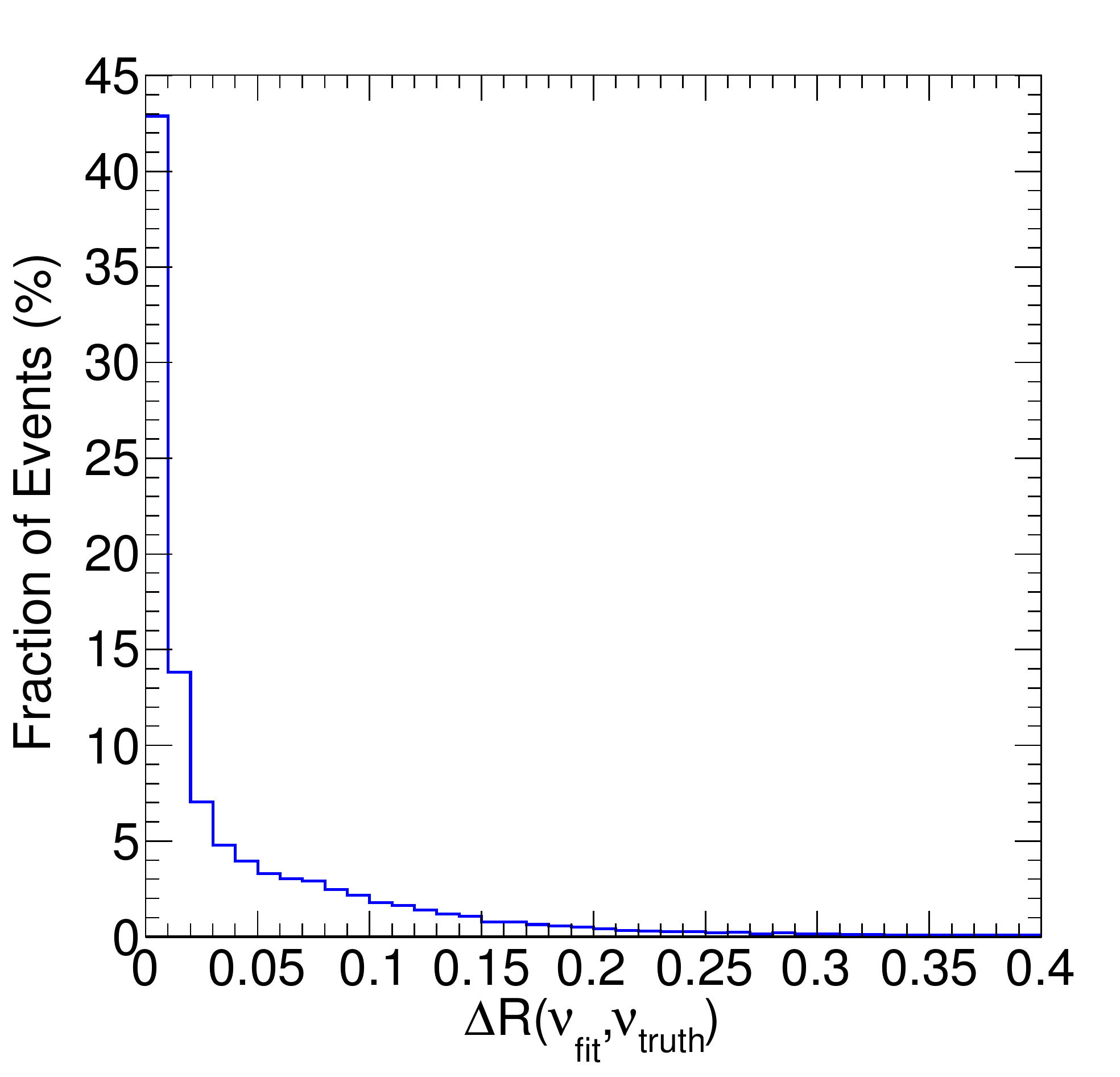}\hspace{2mm}
\put(-45, 65){\textbf{(a)}}
\includegraphics[width=0.36\textwidth]{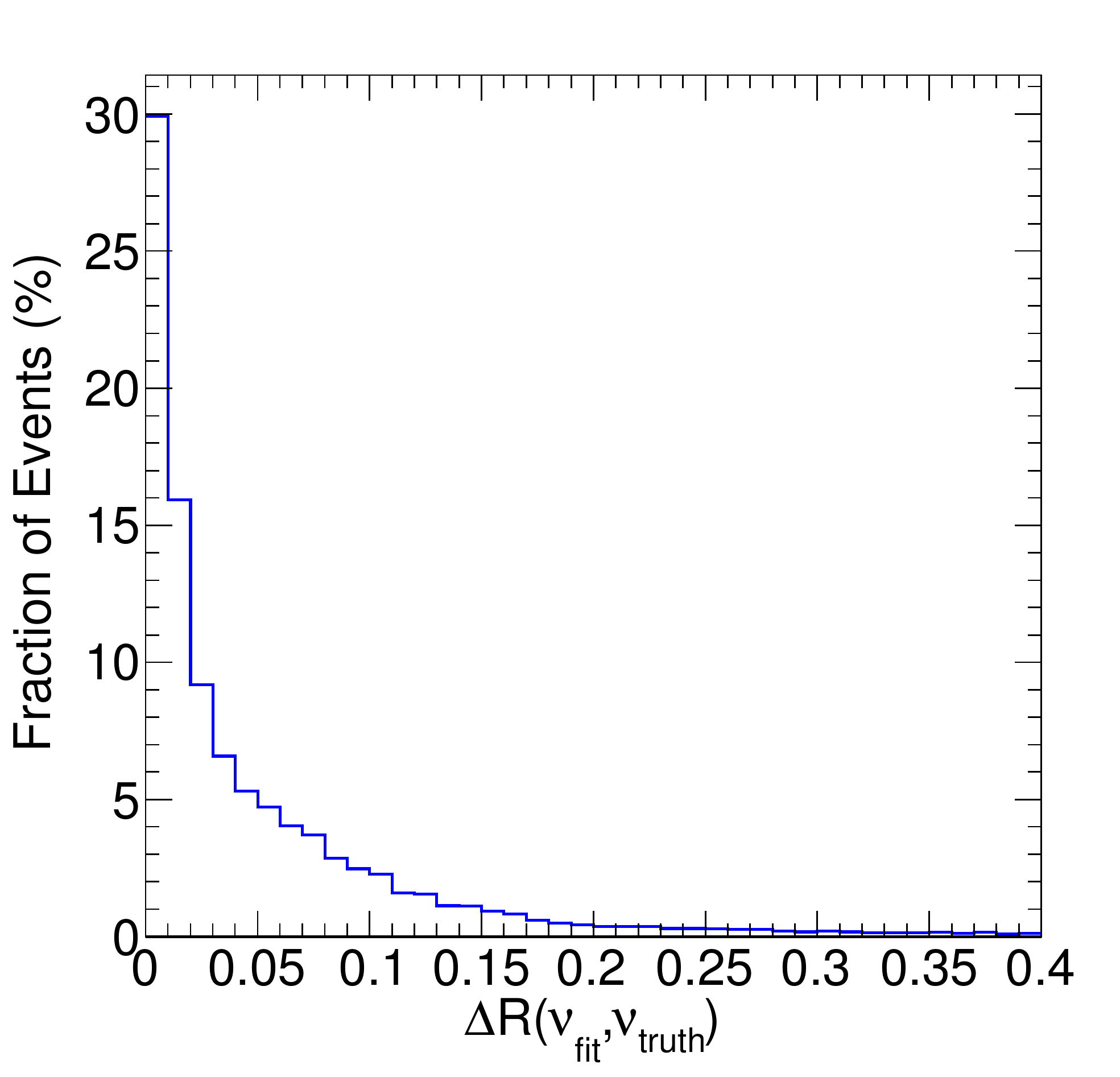}\hspace{2mm}
\put(-45, 65){\textbf{(b)}} \\
\includegraphics[width=0.36\textwidth]{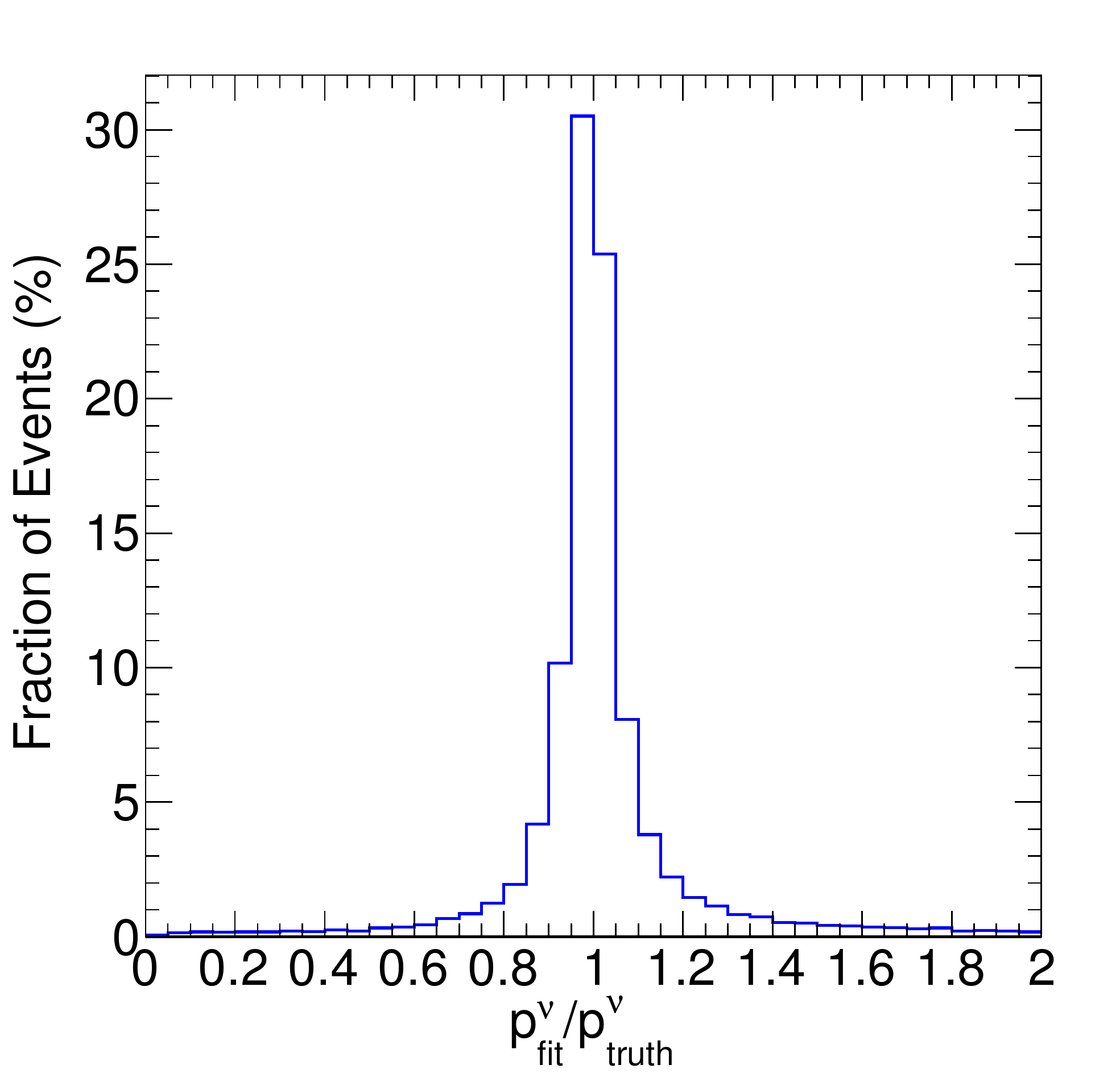}\hspace{2mm}
\put(-45, 65){\textbf{(c)}}
\includegraphics[width=0.36\textwidth]{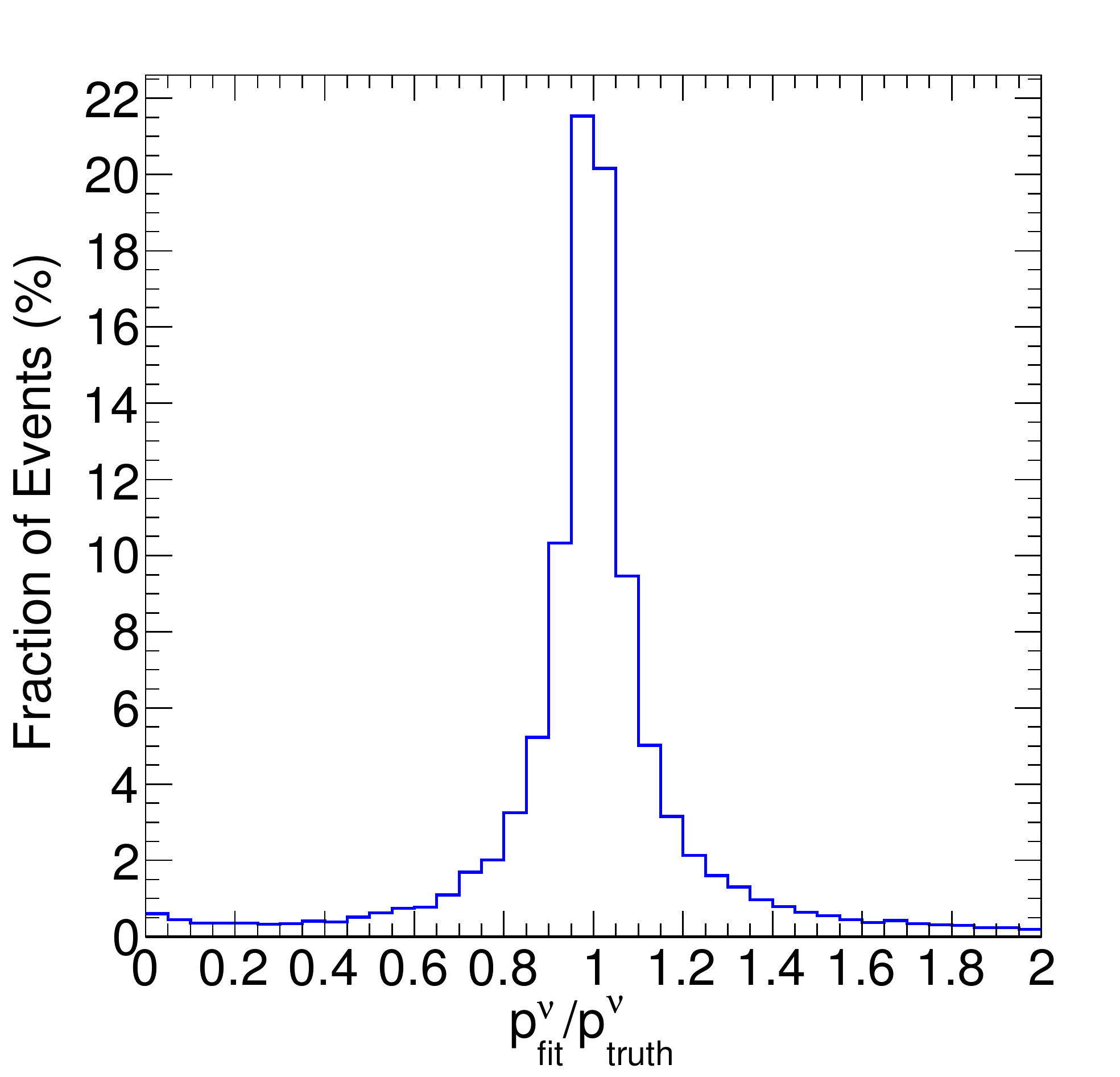}
\put(-45, 65){\textbf{(d)}}
\caption{\label{fig:reco_1} The $\Delta R$ between the reconstructed and true neutrino momenta (a, b), and the ratio of them (c, d), for the $\tau\to\pi\nu$ (a, c) and $\tau\to\rho\nu$ (b, d) decays in the $\pi+\rho$ channel. }
\end{figure*}

In the per-event $\chi^2$ minimization, the ($\eta$, $\phi$) of one neutrino is firstly scanned, from which the magnitudes of the neutrinos' momenta and the direction of the other neutrino can be obtained from the tau mass and recoil constraints in Eq.~\ref{eq:eq4}. Conversely, the scan is repeated from the ($\eta$, $\phi$) of the other neutrino. Finally, a fit using MINUIT~\cite{MINUIT} is performed around the minimal point found by the scans for a better estimation. The $\Delta R$ and momentum ratios between the reconstructed and true neutrinos in the $\pi+\rho$ channel are shown in Fig.~\ref{fig:reco_1}.

\begin{figure*}[tb]
\centering
\includegraphics[width=0.36\textwidth]{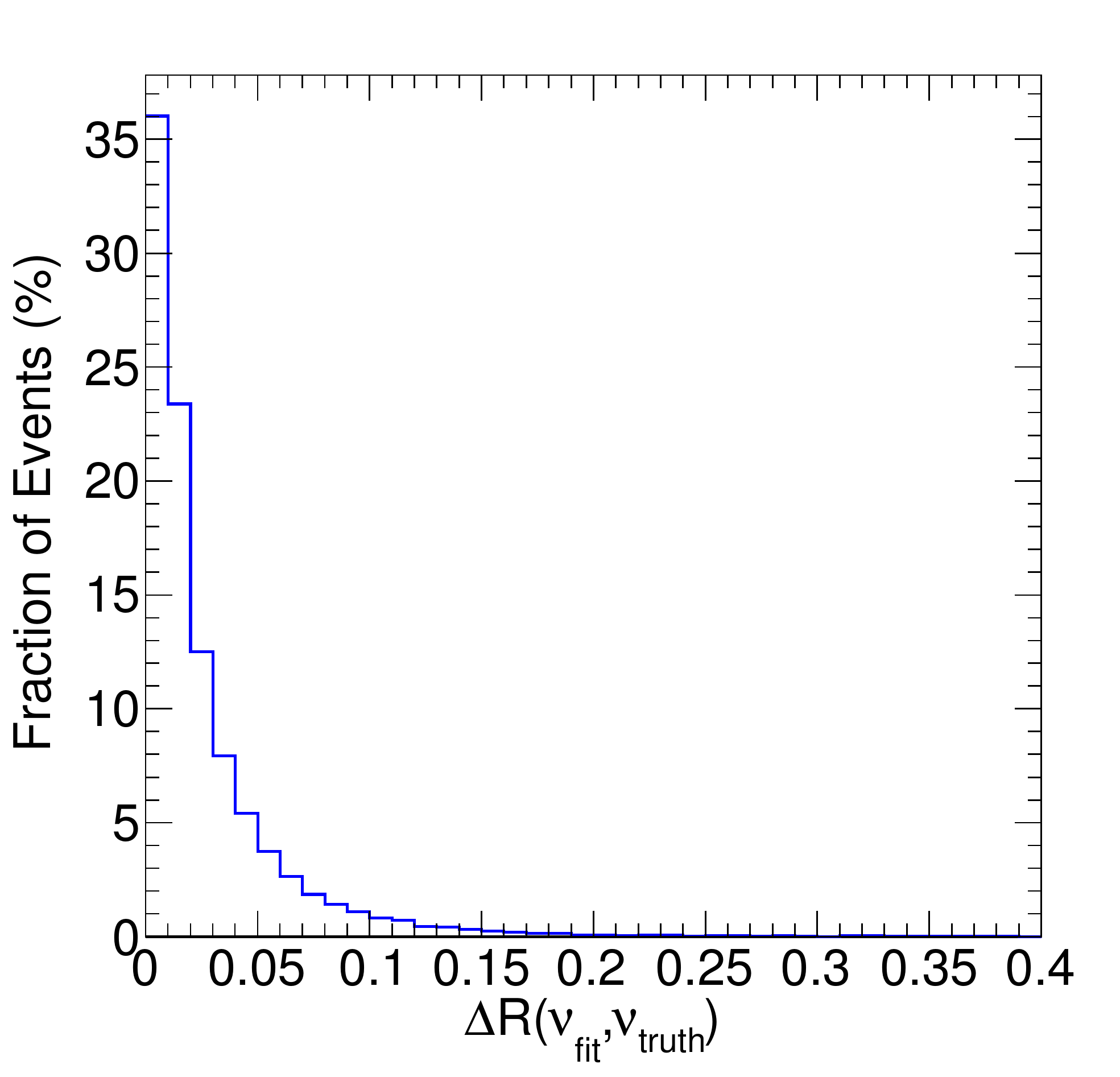}\hspace{2mm}
\put(-45, 70){\textbf{(a)}}
\includegraphics[width=0.36\textwidth]{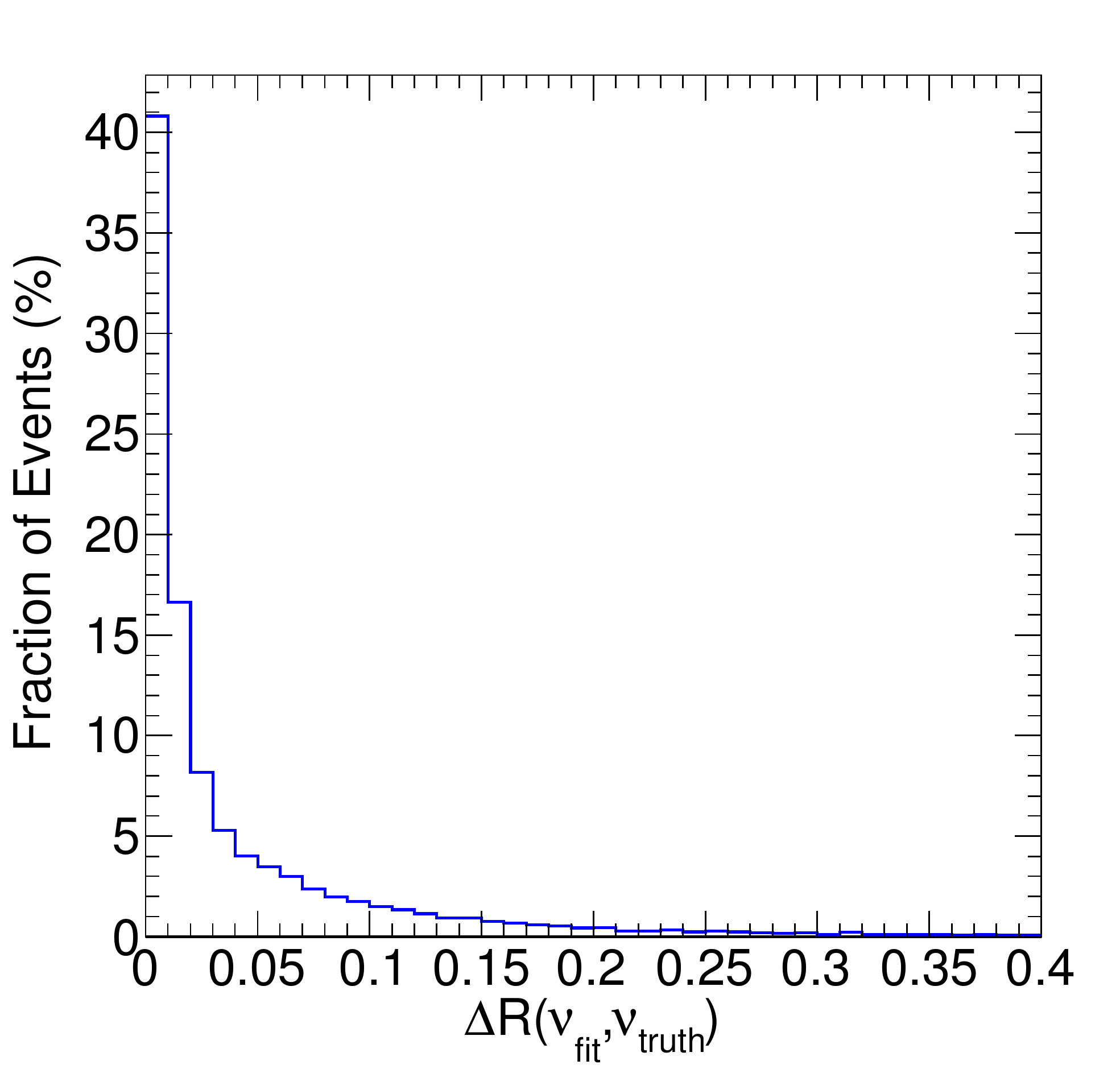}\hspace{2mm}
\put(-45, 70){\textbf{(b)}}\\
\includegraphics[width=0.36\textwidth]{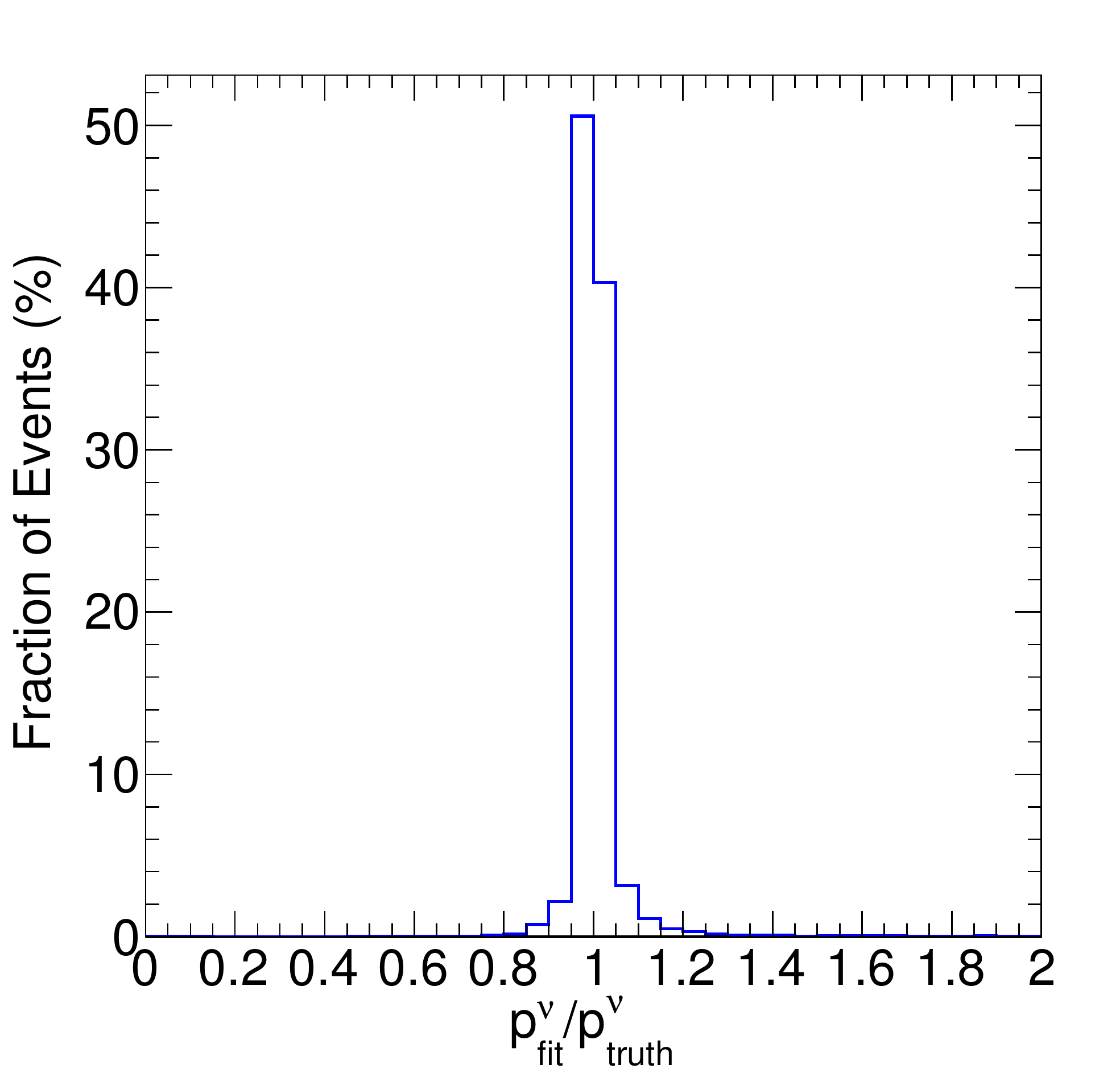}\hspace{2mm}
\put(-45, 70){\textbf{(c)}}
\includegraphics[width=0.36\textwidth]{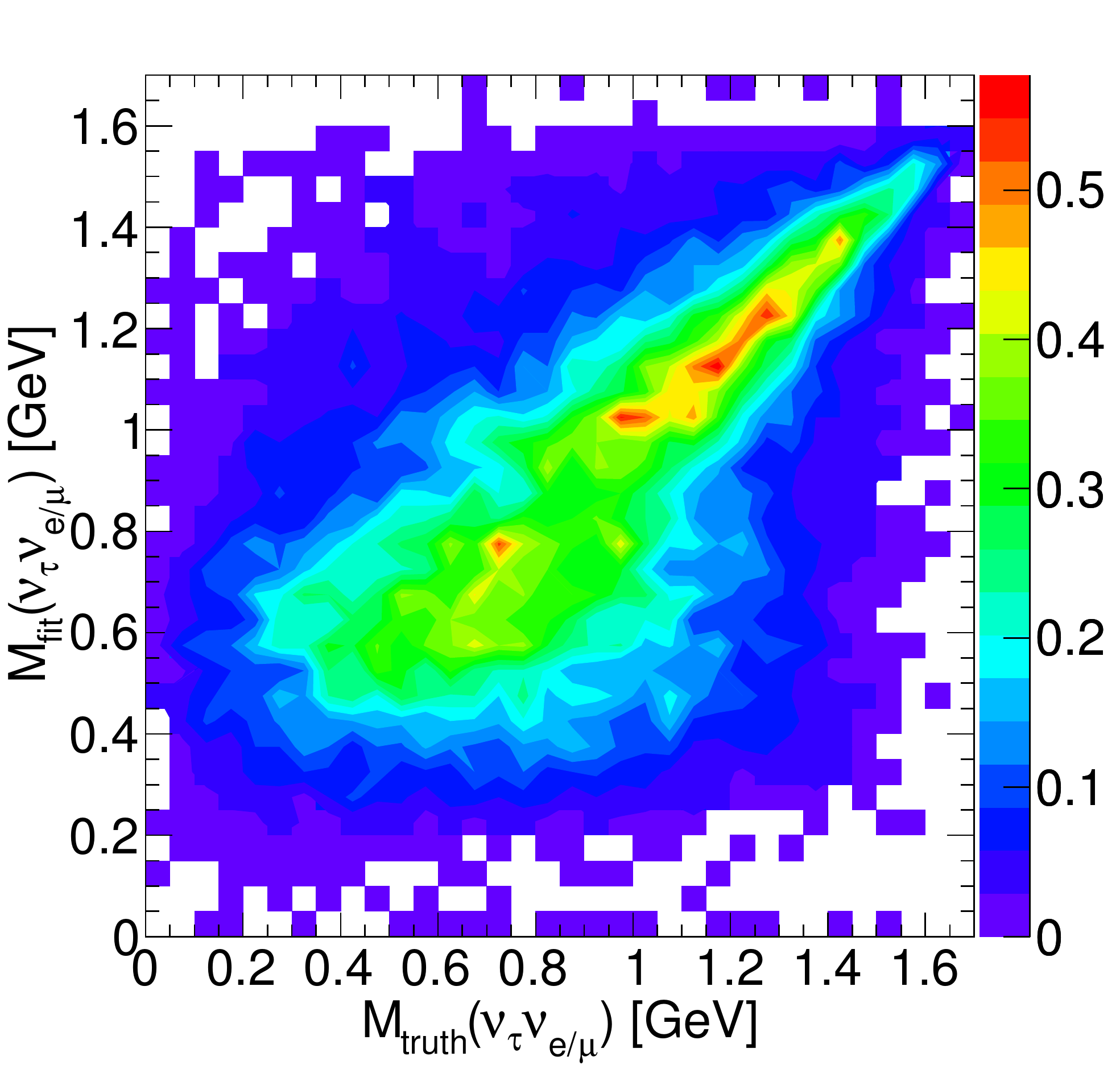}
\put(-45, 70){\textbf{{\color{white}(d)}}}
\caption{\label{fig:reco_2} The $\Delta R$ between the reconstructed and true neutrino (or di-neutrino) momenta for the $\tau\to\ell\nu\nu$ (a) and $\tau\to\pi\nu+\rho\nu$ (b) decays, the ratio of the reconstructed to true di-neutrino momentum for the $\tau\to\ell\nu\nu$ decay (c), and  the 2-D distribution of the true versus reconstructed di-neutrino mass (d), in the $\ell+\pi/\rho$ and $Z\to \ell\ell$ channels. }
\end{figure*}

In the channels with a lepton ($e/\mu$) from tau, the momenta of the two neutrinos from the same tau cannot be fully reconstructed. Instead, the combined four-momenta of them is reconstructed, with the di-neutrino mass as an extra degree of freedom (the relative angles between the two neutrinos are ``integrated'' out). Although there are eight constraints in Eq.~\ref{eq:eq4}, it is still difficult to achieve. In this case, another extra term, $-2 \ln \mathcal{P}(\Delta R, m_{\text{mis}})$, is added. It is the joint probability distribution of $\Delta R$ between the lepton and di-neutrino momenta, and $m_{\text{mis}}$ the di-neutrino invariant mass.  This function is obtained depending on the tau momentum from the Monte Carlo simulation, which has been used in the di-tau MMC mass by ATLAS~\cite{MMC}. The quality of the neutrino momentum reconstruction in the $\ell+\pi/\rho$ and $Z\to \ell\ell$ channels are shown in Fig.~\ref{fig:reco_2}. Due to $m_{\text{mis}}$ and worse Higgs recoil momentum resolutions, the $\ell+\pi/\rho$ channels with $Z\to jj$ have limited reconstruction precision, and are thus excluded from the CP sensitivity test. 

\section{Measurement of the CP mixing angle of \texorpdfstring{$h\tau\tau$}{htautau} at \texorpdfstring{$e^+e^-$}{ee} collider }
\label{sec:MC}
\begin{table}[bt]
\centering
\begin{tabular}{c|c} \hline
 $Z\to \ell\ell$ & $Z\to jj$ \\ \hline
 $m_Z>70$ GeV & $m_Z<105$ GeV \\ \hline
 $m_h^{\text{RC}}>120$ GeV & $m_h^{\text{RC}}>110$ GeV \\ \hline
 $m_{h,\text{fit}}^{\text{RC}}>122$ GeV & 80 GeV$<m_Z^{\text{fit}}<$100 GeV \\ \hline
  \multicolumn{2}{c}{ 120 GeV$<m_h<$130 GeV } \\ \hline
  \multicolumn{2}{c}{ 1.5 GeV$<m_\tau <$2.0 GeV } \\ \hline
  \multicolumn{2}{c}{ $m_\rho>$0.3 GeV (for channels with $\rho$) }\\ \hline
\end{tabular}
\caption{Further kinematic cuts applied to enhance the signal significance.}
\label{tab:tab3}
\end{table}

\begin{figure*}[tb]
\centering
\includegraphics[width=0.32\textwidth]{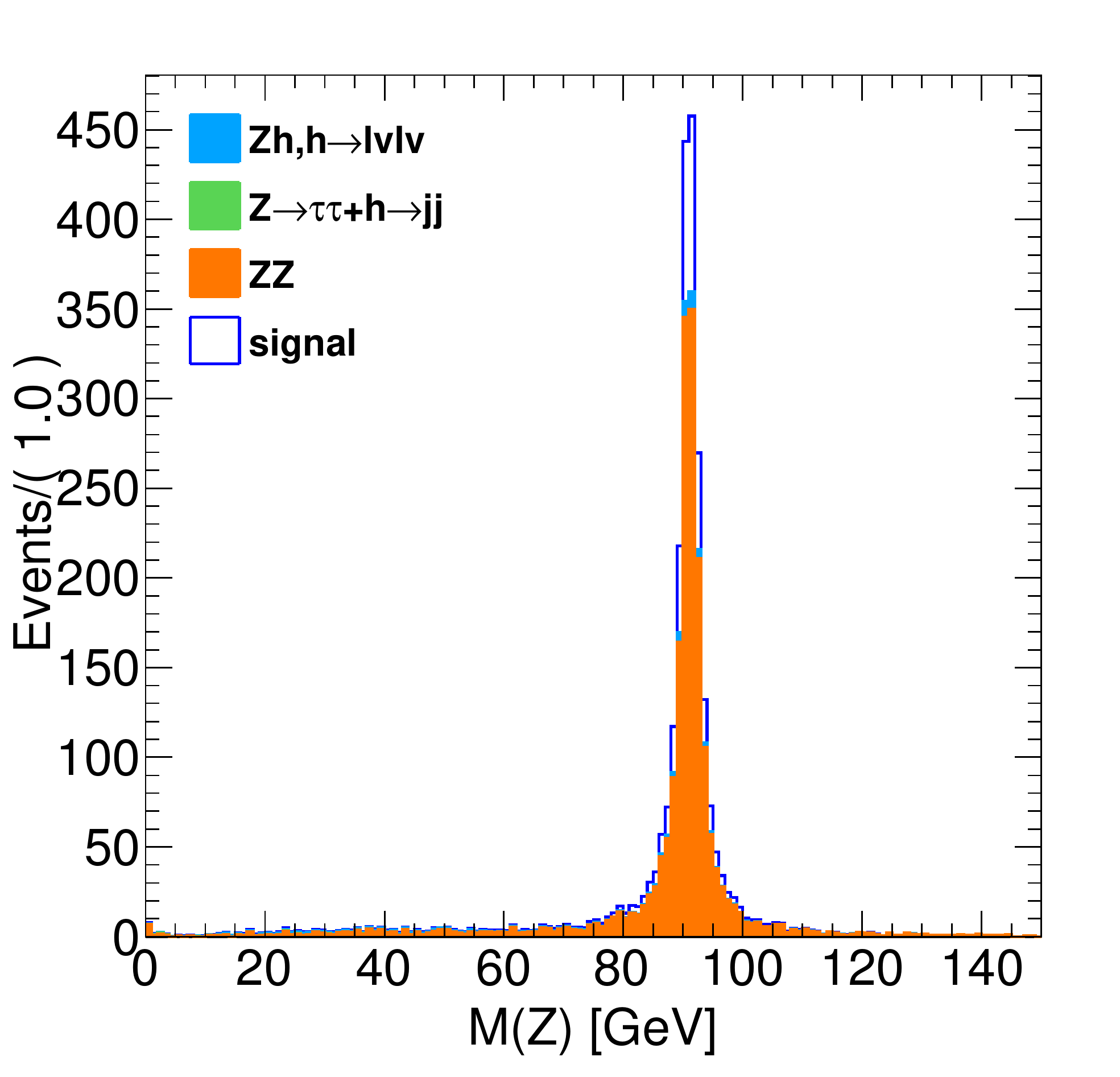}\hspace{2mm}
\put(-45, 65){\textbf{(a1)}}
\includegraphics[width=0.32\textwidth]{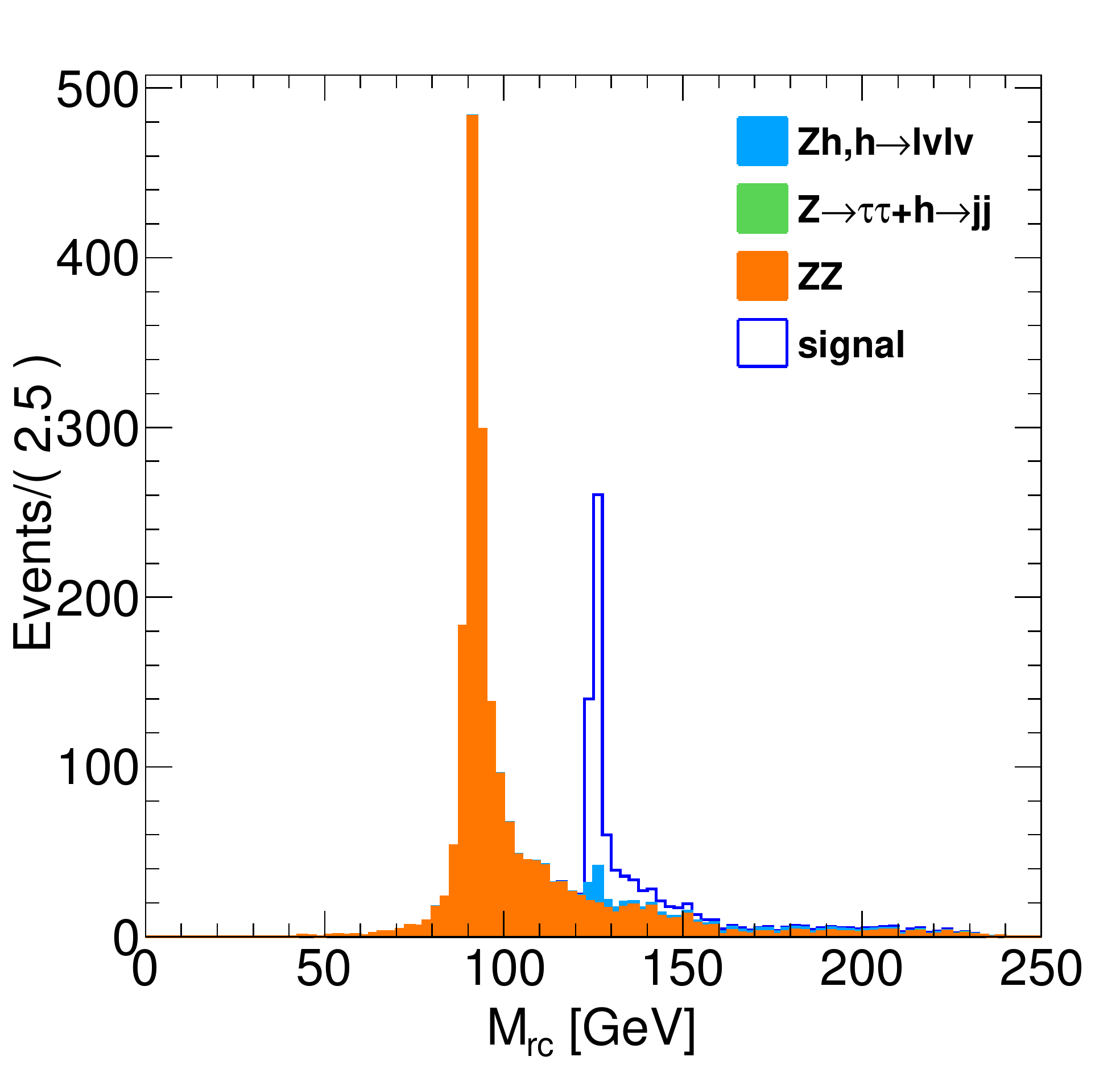}\hspace{2mm}
\put(-45, 65){\textbf{(b1)}}
\includegraphics[width=0.32\textwidth]{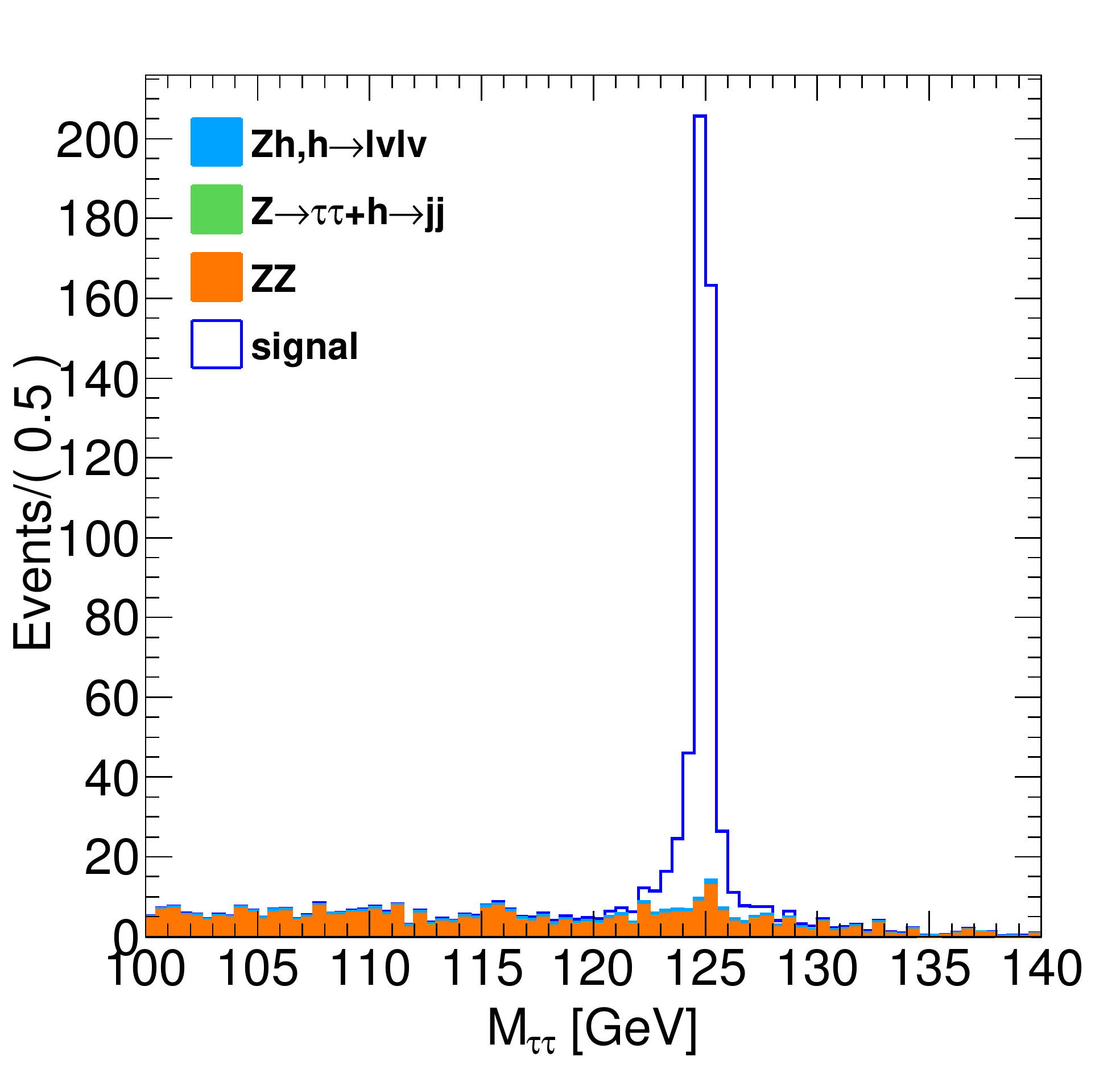}
\put(-45, 65){\textbf{(c1)}}\\
\includegraphics[width=0.32\textwidth]{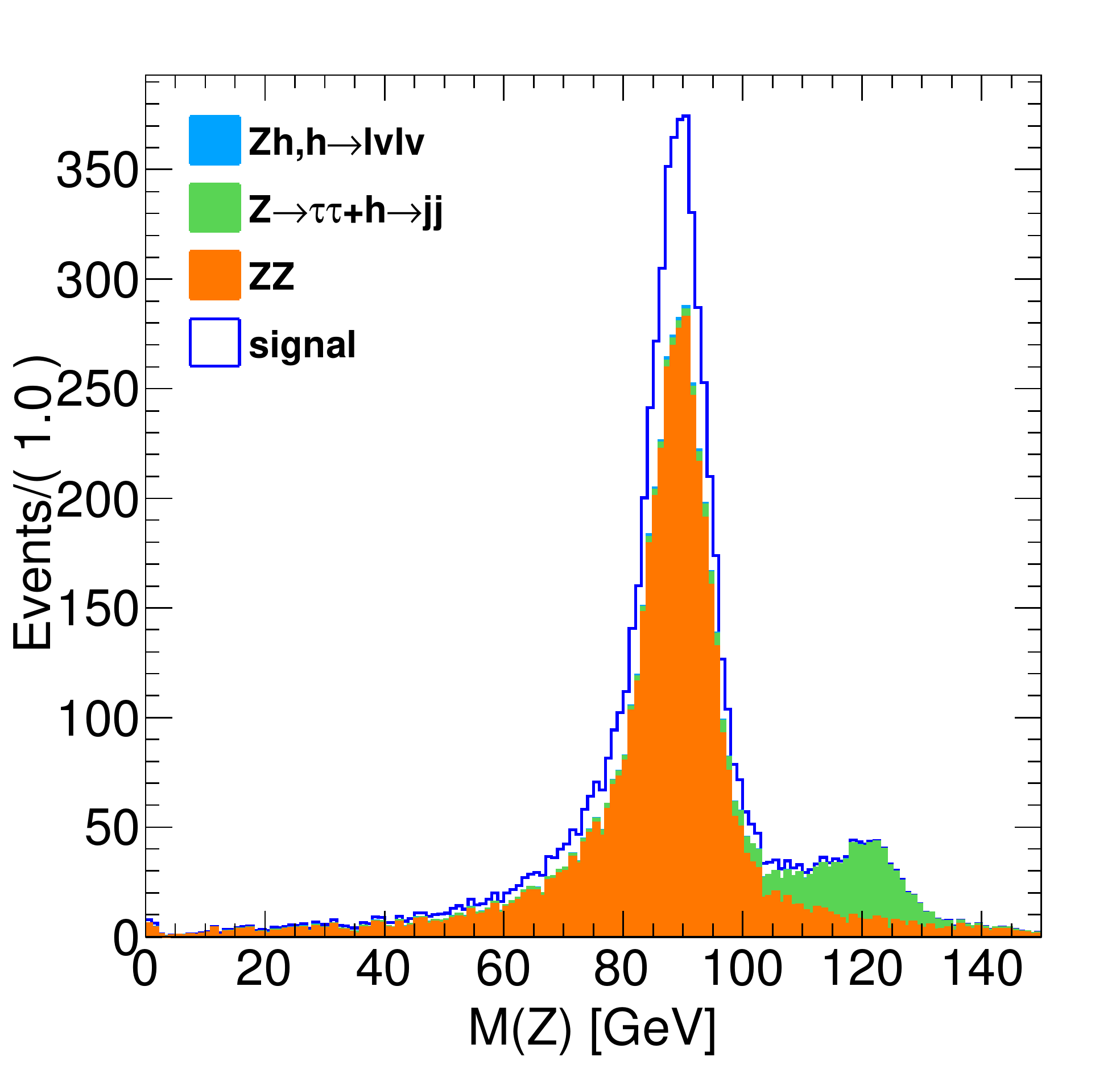}\hspace{2mm}
\put(-45, 65){\textbf{(a2)}}
\includegraphics[width=0.32\textwidth]{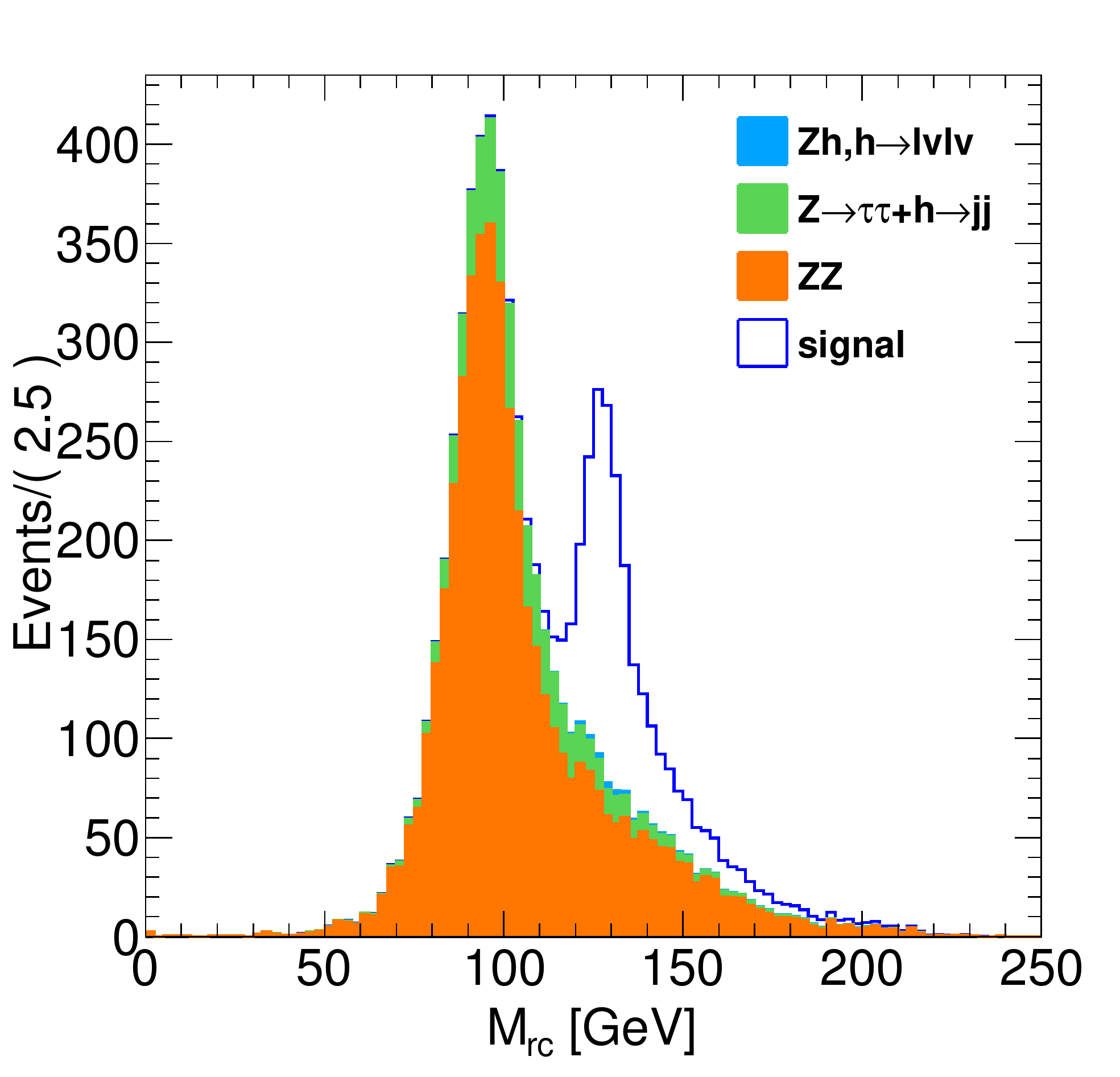}\hspace{2mm}
\put(-45, 65){\textbf{(b2)}}
\includegraphics[width=0.32\textwidth]{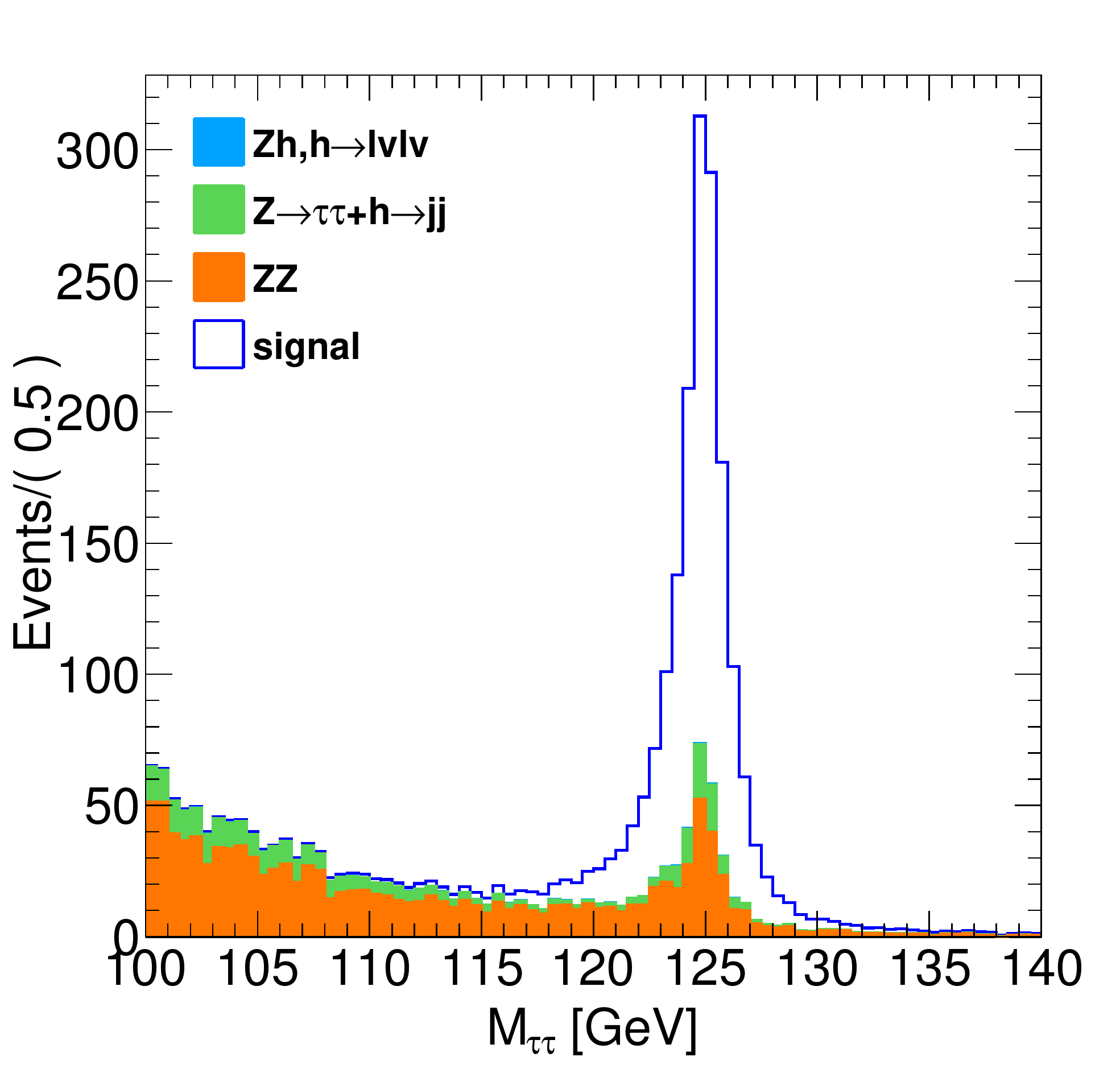}
\put(-45, 65){\textbf{(c2)}}
\caption{\label{fig:cuts} The distributions of $m_Z$ (a1, a2), $m_h^{\text{RC}}$ (b1, b2) and $m_h$ (c1, c2) for the $Z\to \ell\ell$ (a1, b1, c1) and $Z\to jj$ (a2, b2, c2) decays with the three tau decay modes considered and assuming 5 $ab^{-1}$ of data. Note that the $\ell+\pi/\rho$ and $Z\to jj$ channels are excluded. }
\end{figure*}

The simulation details and the basic reconstruction are presented in Sect.~\ref{simu_reco}. To suppress the background, the cuts listed in Tab.~\ref{tab:tab3} are further applied to achieve the best signal significance. The $m_h^{\text{RC}}$ is calculated from the $Z$ recoil, $m_{h,\text{fit}}^{\text{RC}}$ and $m_Z^{\text{fit}}$ are obtained from per-event fit, and $m_h$, $m_\tau$ and $m_\rho$ are calculated from the di-tau side. Fig.~\ref{fig:cuts} shows the distributions of $m_Z$, $m_h^{\text{RC}}$ and $m_h$ for the $Z\to \ell\ell$ and $Z\to jj$ decays before the cuts are applied. With 5 $ab^{-1}$ of $e^+e^-$ collision data at $E_{\text{CM}}=250$ GeV, the expected numbers of events entering the CP sensitivity test are listed in Tab.~\ref{tab:tab4}. Note that the whole per-event fitting procedure not only helps to determine the momentum of neutrino and also provides better resolutions for these mass observables and thus helps to further suppress the backgrounds. 

\begin{table}[tb]
\centering
\begin{tabular}{c|ccccc} \hline
  & $\ell+\pi$ & $\ell+\rho$ & $\pi+\pi$ & $\pi+\rho$ & $\rho+\rho$ \\ \hline
signal & 112.4 & 194.8 & 147.2 & 541.6 & 523.2 \\ \hline
background & 9.5 & 12.6 & 15.5 & 46.7 & 48.6 \\ \hline
\end{tabular}
\caption{The expected numbers of signal and background events with 5 $ab^{-1}$ of data in each channel after the cuts in Tab.~\ref{tab:tab3} with $Z\to \ell\ell,jj$ combined. Note that the $\ell+\pi/\rho$ and $Z\to jj$ channels are excluded.}
\label{tab:tab4}
\end{table}

\subsection{Matrix element-based measurement of CP}
\label{cp_fit}

With the fully reconstructed momentum of final states from the Higgs decay, a method based on the matrix elements could be used to probe the CP mixing angle.  From Eq.~\ref{eq:eq1}, the matrix element squared with $\phi$ being the CP mixing angle can be expressed as:
\begin{eqnarray}
&& \qquad |\mathcal{M}|^2 \propto  A + B \cos(2\phi) + C \sin(2\phi), \nonumber \\
		&&		\propto  I_1\cos^2(\phi) + I_2\sin(\phi)\cos(\phi) + I_3\sin^2(\phi), \
\end{eqnarray}
where $I_1 = A+B$, $I_2 = 2C$ and $I_3 = A-B$. For example, in the $\pi+\pi$ channel, the coefficients have relatively simple expressions:
\begin{eqnarray}
\label{equ:MEpipi}
A_{\pi\pi} &=& \frac{m_{\tau}^4}{8}((m_h^2 - 2m_{\tau}^2)(m_{\tau}^2 - m_{\pi}^2)^2 - 8m_{\tau}^4(p_{\pi^{-}} \cdot  p_{\pi^{+}})) \nonumber \\
B_{\pi\pi} &=& -\frac{m_{\tau}^6}{4}((m_{\pi}^2 - m_{\tau}^2)^2 - 2(m_h^2 - 2m_{\tau}^2)(p_{\pi^{-}} \cdot  p_{\pi^{+}}) \nonumber \\
& &+ 4(p_{\pi^{-}} \cdot  p_{\tau^{+}})(p_{\pi^{+}} \cdot  p_{\tau^{-}})) \nonumber \\
C_{\pi\pi} &=& m_{\tau}^6\epsilon^{p_{\pi^{-}}  p_{\pi^{+}}  p_{\tau^{-}}  p_{\tau^{+}}  }
\end{eqnarray}
where $p_{\tau^-}$ and $p_{\tau^+}$ are the momenta of the two taus, which include the information of the reconstructed neutrinos and $\epsilon^{p_1p_2p_3p_4}$ is short for $\epsilon_{\mu\nu\rho\sigma}p_1^\mu p_2^\nu p_3^\rho p_4^\sigma$.

\begin{figure*}[tb]
\centering
\includegraphics[width=0.33\textwidth]{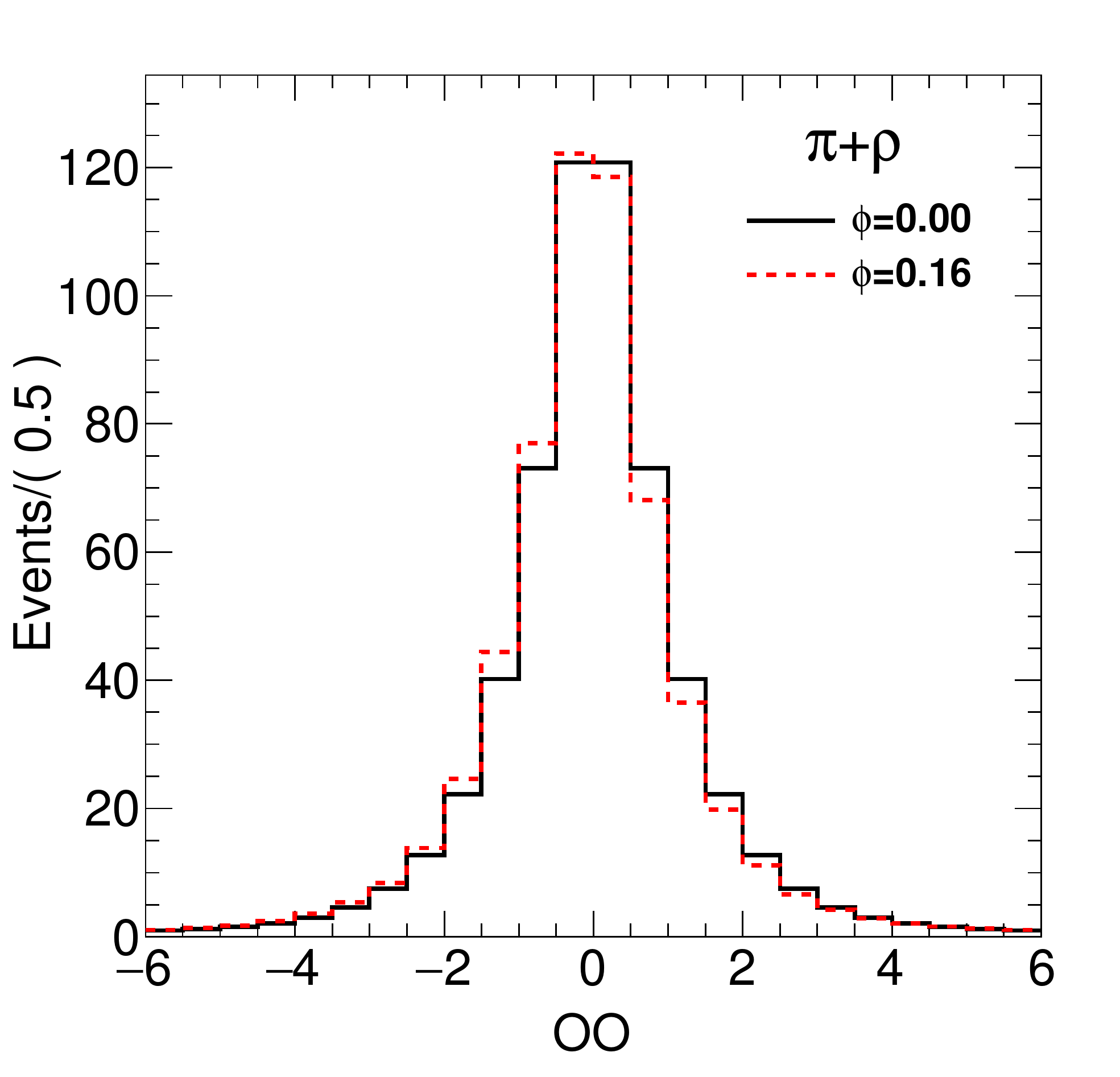}
\put(-60, 100){\textbf{(a)}}
\includegraphics[width=0.33\textwidth]{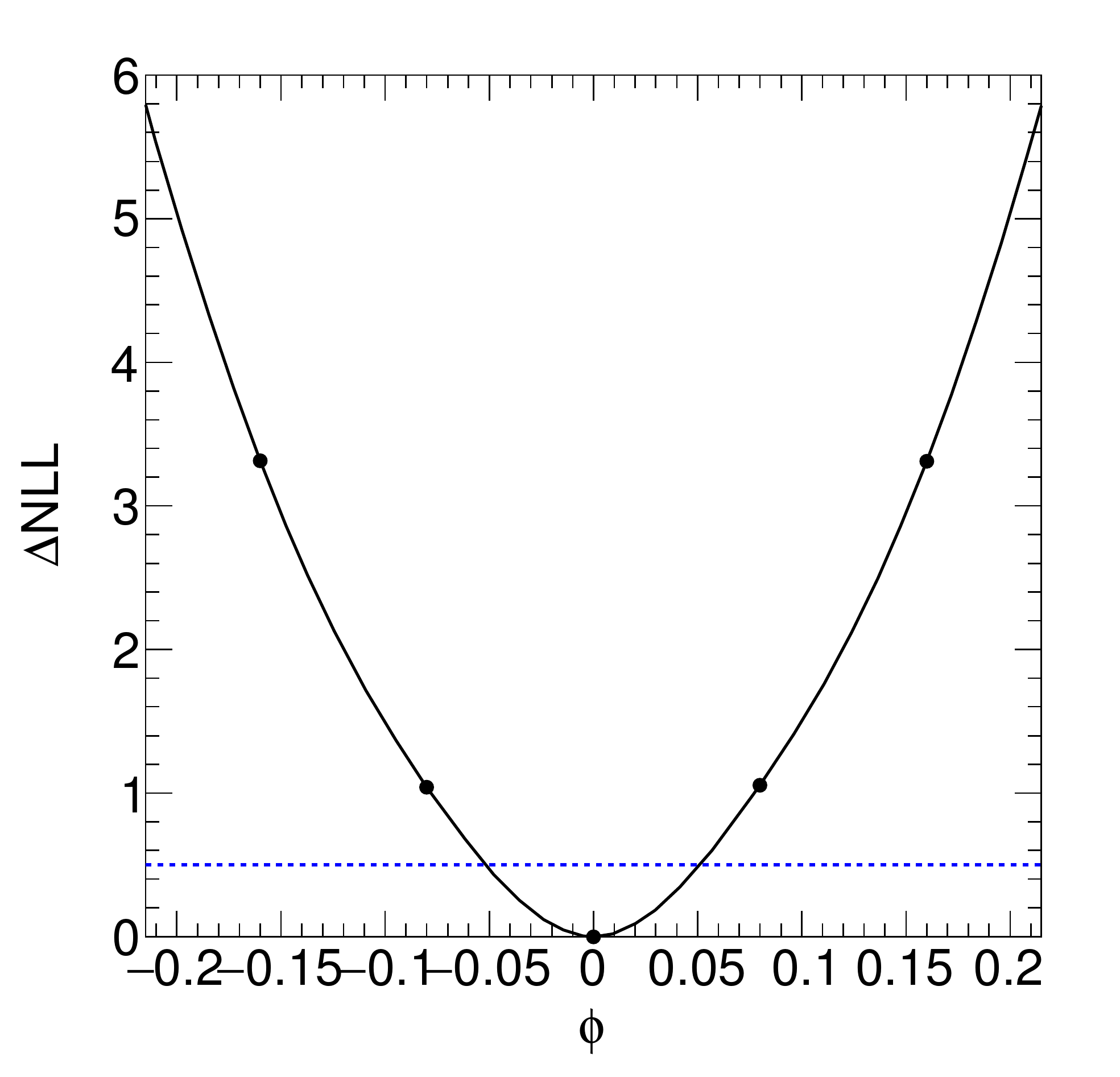}
\put(-60, 100){\textbf{(b)}}
\caption{\label{fig:oo_nll} (a): the expected $OO$ distributions in the $\pi+\rho$ channel with 5 $ab^{-1}$ of data for two different CP mixing angles. (b): the $\Delta$NLL as a function of the CP mixing angle $\phi$ calculated from expected zero-CP events at 5 $ab^{-1}$, with all five channels in Tab.~\ref{tab:tab4} combined. The $1\sigma$ interval bounds are the intersections between the $\Delta$NLL curve and the blue dashed line. }
\end{figure*}

The full expression of the coefficients for the other decay mode combinations can be found in Appendix~\ref{append} where we also list the effective Lagrangian used for tau decay. For channels involving leptonic decays of tau, a phase space integral is performed on the internal degrees of freedom between the two neutrinos, at the cost of a bit loss of sensitivity to CP. In the $\rho+\rho$ and $\pi+\rho$ channels, the neutral pion(s) is assigned to the corresponding decay chain without any ambiguity, since the neutral pion from different decay chain is highly collimated with the original tau.  An optimal obsevable~\cite{oo1,oo2}, defined as $OO=I_2/I_1$ for each event, is used to distinguish signals with different CP mixing angles. For signals with a positive (negative) $\phi$, the mean of the $OO$ distribution will be shifted to negative (positive) values, as shown in Fig.~\ref{fig:oo_nll}(a). 

Template probability density functions (PDF) for different CP angles are first obtained from the simulations. A binned likelihood function ($\mathcal{L}$) is then calculated from the PDF as a function of the CP mixing angle $\phi$. 
The best-fit CP angle is found with the least negative logarithm of the likelihood (NLL), and the $1\sigma$ confidence interval is obtained by finding the angles whose NLLs are 0.5 higher than the minimum NLL. Fig.~\ref{fig:oo_nll}(b) shows the expected $\Delta$NLL as a function of the CP angle $\phi$, with all five channels in Tab.~\ref{tab:tab4} combined. With 5 $ab^{-1}$ of $e^+e^-$ collision data (can be achieved by the CEPC~\cite{CEPC-CDR}), a precision of $0.05$ radians can be reached. On the other hand, if the integrated luminosity is 2 $ab^{-1}$ (can be achieved by the ILC~\cite{Behnke:2013xla}), a precision of about $0.09$ can be reached.

\begin{figure*}[tb]
\centering
\includegraphics[width=0.33\textwidth]{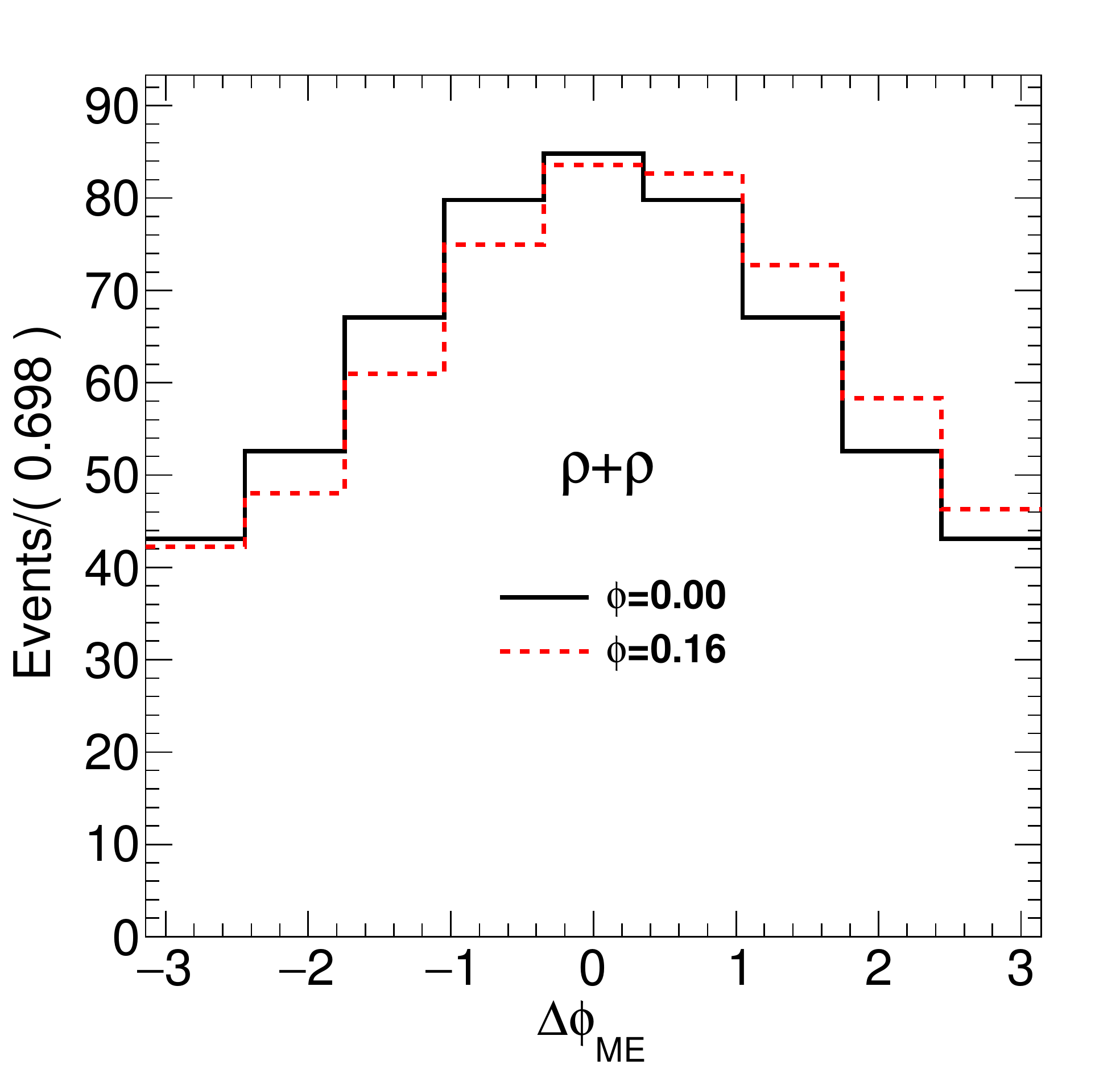}
\put(-40, 50){\textbf{(a)}}
\includegraphics[width=0.33\textwidth]{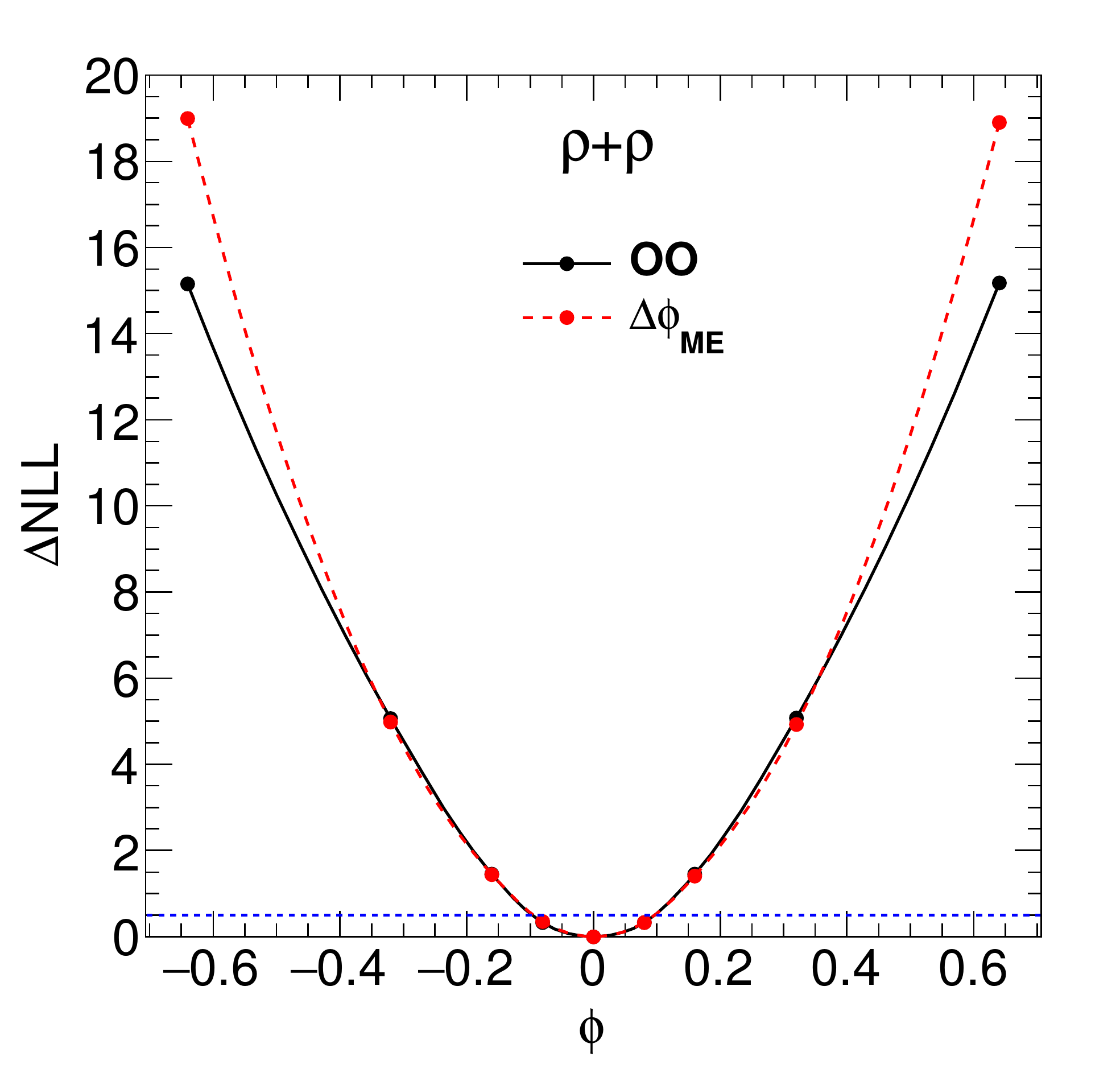}
\put(-40, 50){\textbf{(b)}}
\caption{\label{fig:me_nll}(a): the expected $\Delta\phi_{ME}$ distributions in the $\rho+\rho$ channel with 5 $ab^{-1}$ of data for two different CP mixing angles. (b): the $\Delta$NLL as a function of the CP mixing angle $\phi$ calculated from expected zero-CP events at 5 $ab^{-1}$ for $\Delta\phi_{ME}$ and OO in $\rho\rho$ channel.}
\end{figure*}

Actually, one can also built an angle $\Delta\phi_{ME}$ ($-\pi\leq\Delta\phi_{ME}\leq\pi$) to probe the CP mixing angle which satisfies:
\begin{eqnarray}
&&\cos(\Delta\phi_{ME}) = \frac{B}{\sqrt{B^2+C^2}}, \nonumber \\
&&\sin(\Delta\phi_{ME}) = \frac{C}{\sqrt{B^2+C^2}}.
\end{eqnarray}
With this definition, the matrix element (which also determines the distribution of $\Delta\phi_{ME}$) becomes
\begin{eqnarray}
|\mathcal{M}|^2 \propto A + \sqrt{B^2+C^2}\cos(\Delta\phi_{ME}-2\phi)
\end{eqnarray}
 The distribution of $\Delta\phi_{ME}$ is presented in Fig.~\ref{fig:me_nll}(a) with two different choice of CP mixing angle $\phi$. $\Delta\phi_{ME}$ makes full use of the information inside the matrix element and performs better when $\phi$ is large. Fig.~\ref{fig:me_nll}(b) shows the NLL comparison between OO and $\Delta\phi_{ME}$, we find that when $\phi$ is large, $\Delta\phi_{ME}$ is better than OO (as in the OO construction, the term $I_3$ is omitted). But in the small $\phi$ region, the two methods give quite similar results.

Further, the $\rho+\rho$ channel is used to compare the sensitivities of the $\Delta\phi_{\text{IP}}$ method (using the impact parameters to approximately reconstruct the tau decay planes), the $\Delta\phi_{\text{CP}}$ method (using the visible $\rho$ meson decay products to reconstruct the CP sensitive angle $\Delta\phi_{\text{CP}}$) and the $OO$ method, the details of the definitions for $\Delta\phi_{\text{IP}}$ and $\Delta\phi_{\text{CP}}$ are given in Appendix~\ref{appPhi}.  Figure~\ref{fig:comp}(a, b) shows the distributions of $\Delta\phi_{\text{IP}}$, $\Delta\phi_{\text{CP}}$ in this channel. It is seen that the IP method have much worse separation power between CP even and CP mixed states than the $\Delta\phi_{\text{CP}}$ method. Figure~\ref{fig:comp}(c) shows the $\Delta$NLL variations with the $\Delta\phi_{\text{CP}}$ and the OO, from which one can read off that the latter is about $35\%$ better than the former.

\begin{figure*}[tb]
\centering
\includegraphics[width=0.32\textwidth]{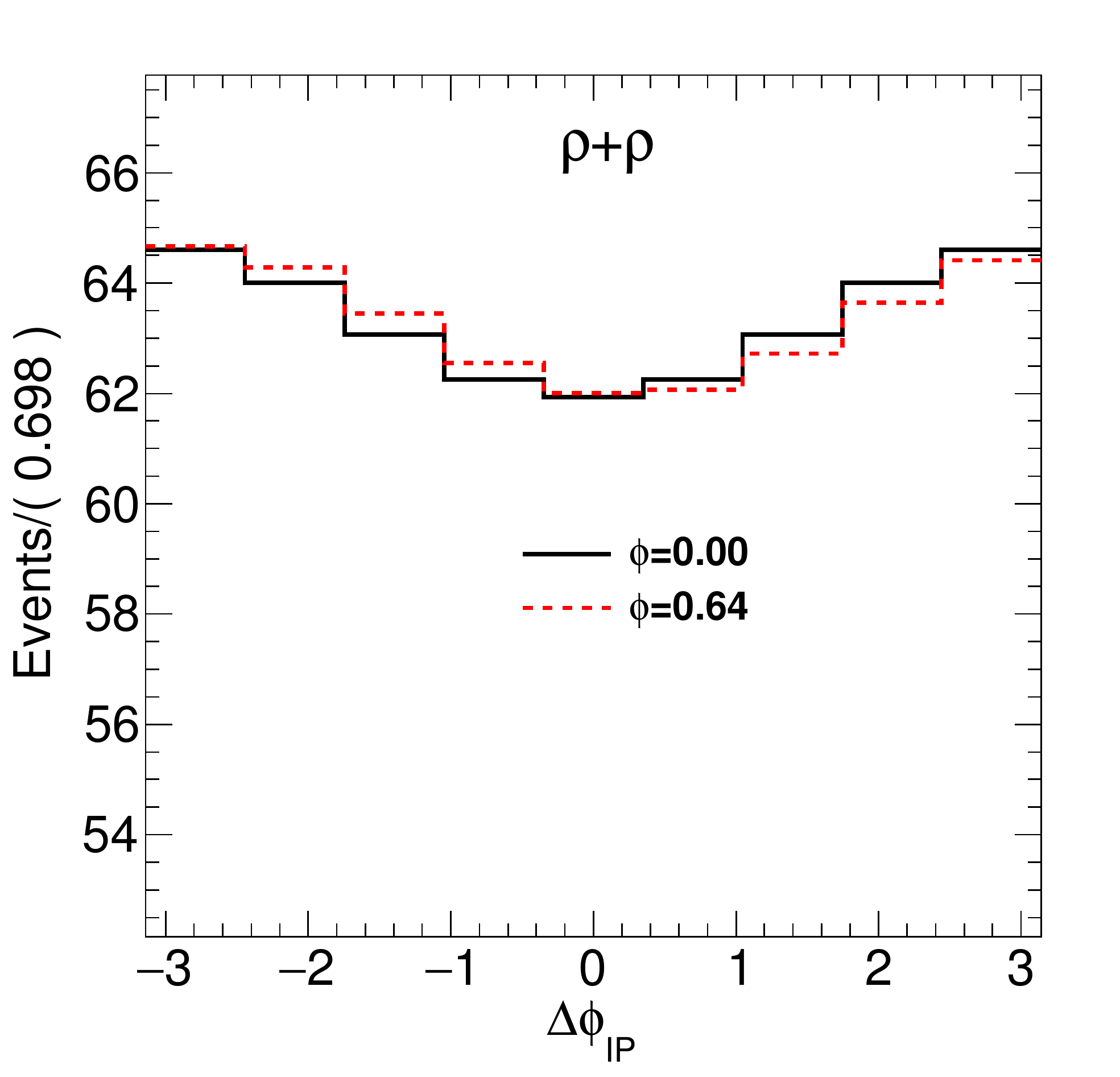}\hspace{2mm}
\put(-70, 30){\textbf{(a)}}
\includegraphics[width=0.32\textwidth]{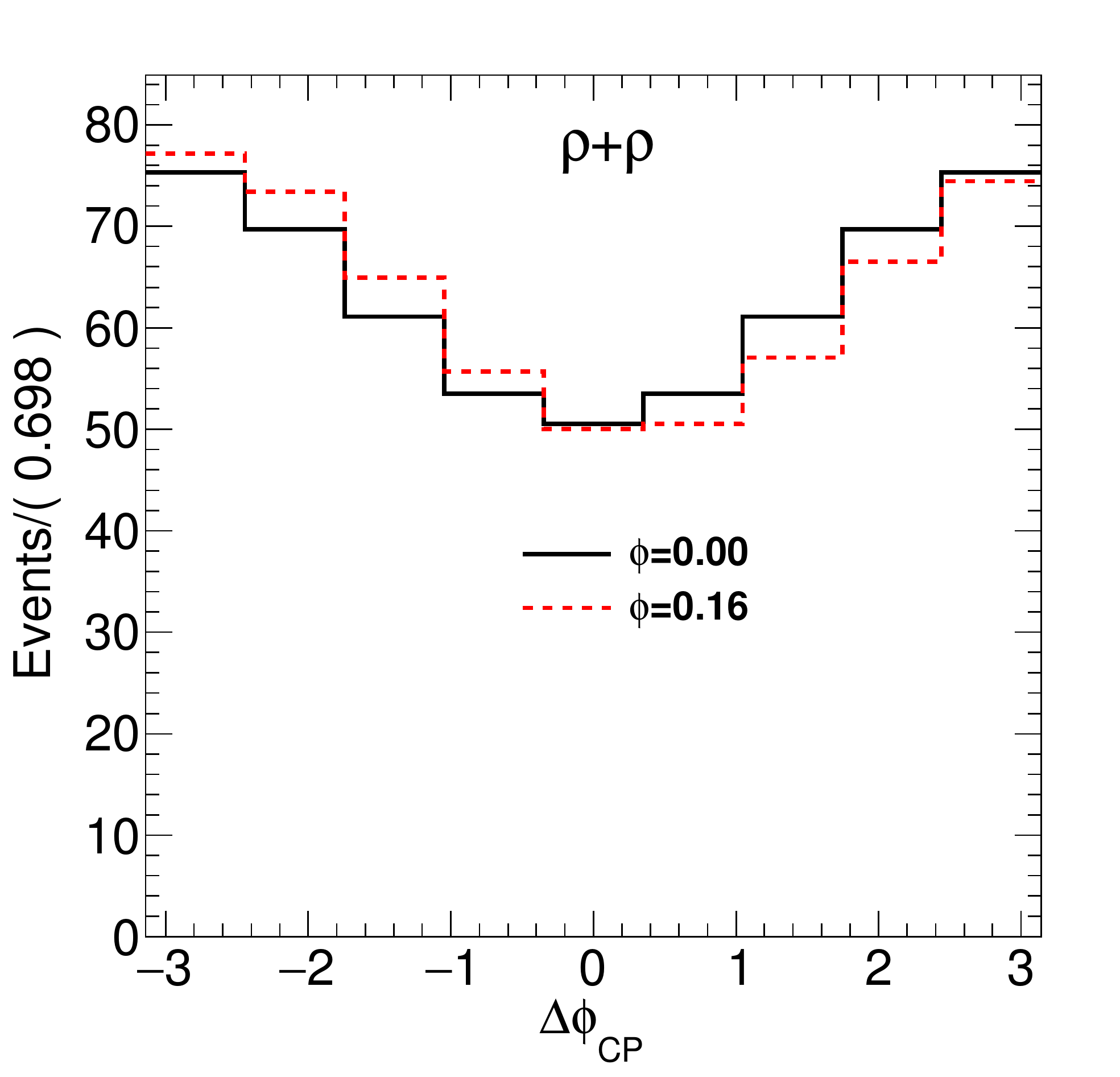}\hspace{2mm}
\put(-70, 30){\textbf{(b)}}
\includegraphics[width=0.32\textwidth]{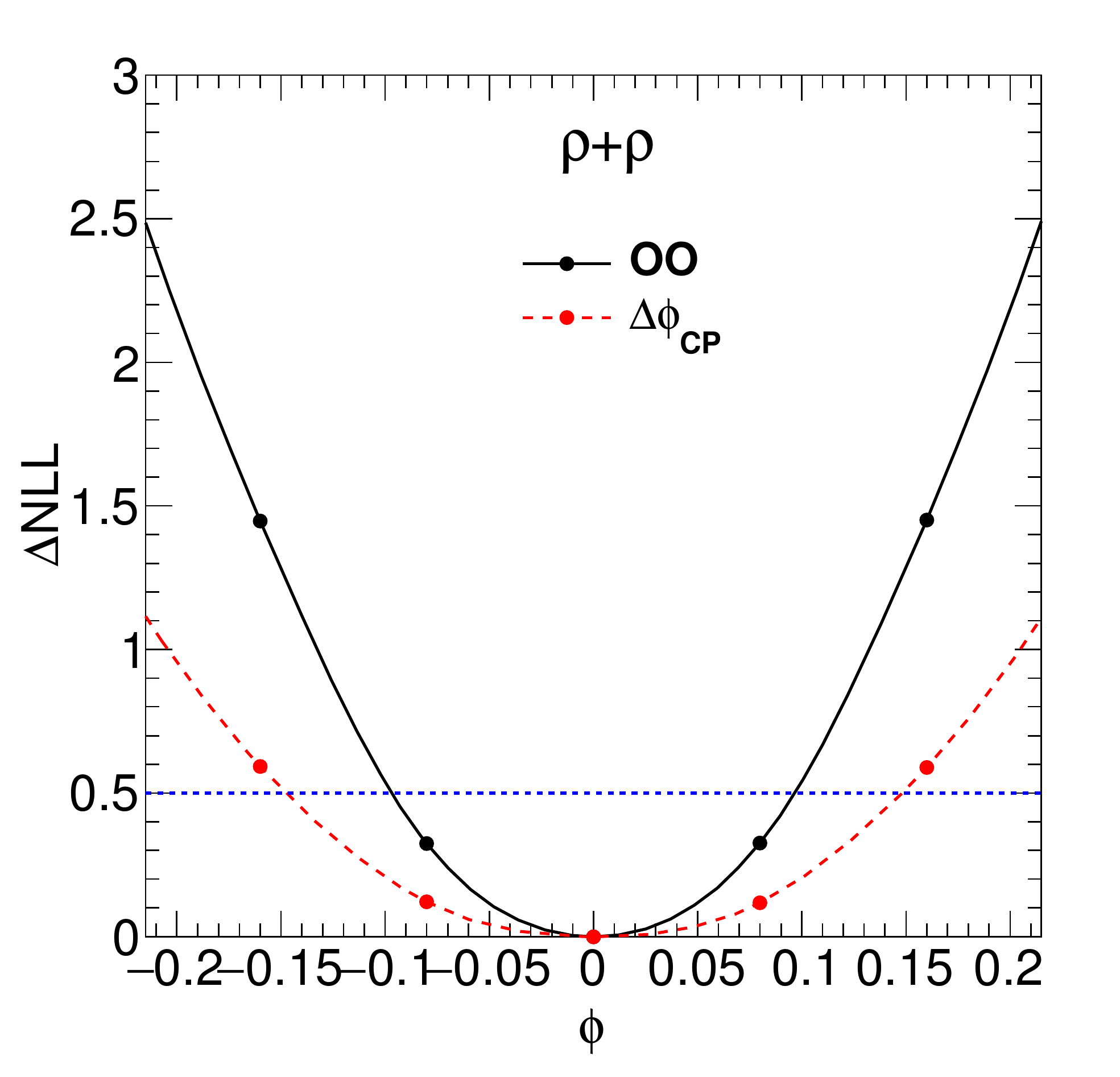}
\put(-70, 50){\textbf{(c)}}
\caption{\label{fig:comp} The distributions of $\Delta\phi_{\text{IP}}$ (a) and $\Delta\phi_{\text{CP}}$ (b) in the $\rho+\rho$ channel with 5 $ab^{-1}$ of data, using the impact parameter and $\Delta\phi_{\text{CP}}$ method respectively. The $\Delta$NLL variations with the $\Delta\phi_{\text{CP}}$ and the OO method are compared in (c) with the same channel. }
\end{figure*}

\section{Conclusion}
\label{conclusion}
Understanding the CP property of the Higgs is one of the primary goals in future $e^+e^-$ colliders or Higgs factory. If the Higgs is found to be a CP mixture, a door to matter-antimatter imbalance and New Physics would be opened. The $H\to\tau\tau$ decay has a unique position in the Higgs CP search, as the CP odd contribution can enter the Lagrangian at the tree level. 

In this paper, we proposed a new method utilizing the impact parameters and the on-shell conditions to fully reconstruct the momentum of the neutrinos from the tau decay (in the leptonic decay channel, the sum of the two neutrinos). With fully reconstructed the momenta of final state particles from the tau decays using a global likelihood scan and fit, the Higgs CP information can be best estimated based on a matrix element method. In the reconstruction, the resolution degradation due to ISR photons can be reduced by a pre-fit of the Higgs' momentum from the $Z$ recoil, which takes into account the Higgs mass and the special $H\to\tau\tau$ decay topology. 

We have performed a complete MC simulation taking into account the performance of current LHC detectors for the signal process $e^+e^-\to Zh\to (\ell\ell/jj)(\tau^+\tau^-)$ followed by three main one-prong decay channels of tau and also the corresponding backgrounds. The refined Higgs momentum can further help to reduce the background leaving enough signal rates. By virtue of fully reconstructed neutrinos, the matrix element is calculated event by event for all di-tau decay modes, and matrix element based observables are built (OO and $\Delta\phi_{\text{ME}}$) to probe the CP mixing angle $\phi$ of the $h\tau\tau$ coupling. At larger value of $\phi$, $\Delta\phi_{\text{ME}}$ is definitely better than OO, but for small values of $\phi$, the performances of them are found to be similar.

Template PDFs are obtained from the simulation, and the NLL analysis based on the PDF shows that with 5 $ab^{-1}$ (2 $ab^{-1}$) of data at $E_{\text{CM}}=250$ GeV center of mass energy from $e^+e^-$ collisions, a CP mixing angle can be measured to a precision of $0.05$ ($0.09$) radians, or $2.9^\circ$ ($5.2^\circ$). The comparison with other CP observables ($\Delta\phi_{\text{IP}}$ and $\Delta\phi_{\text{CP}}$) has also been presented, and we find that the CP mixing angle sensitivity reach based on our method is about at least 35\% better than the other methods. 
The method can also be extended to other collider analyses that are sensitive to the neutrino momentum reconstruction.
\appendix
\section{Matrix element expressions for the channels considered in this work}
\label{append}
The effective Lagrangian we used for the calculation of the matrix elements is
\begin{eqnarray}
\mathcal{L} &=& -\frac{y_\tau}{\sqrt{2}} ( \cos\phi\bar{\tau}\tau + i \sin\phi\bar{\tau}\gamma_5\tau) h \nonumber \\
&& + C_\pi(\bar{\tau}\gamma^\mu P_L\nu_\tau\partial_\mu\pi^- + \text{h.c.}) \nonumber \\
&& + C_\rho(\bar{\tau}\gamma^\mu P_L\nu_\tau(\pi^0\partial_\mu\pi^--\pi^-\partial_\mu\pi^0)+\text{h.c.}) \nonumber \\
&& + \sqrt{2}G_F(\bar{\tau}\gamma^\mu P_L\nu_\tau \bar{\nu_\ell}\gamma_\mu P_L\ell + \text{h.c.})
\end{eqnarray} 

For each Channel, the matrix element square has following form:
\begin{eqnarray}
|\mathcal{M}|^2 &\propto & A + B \cos(2\phi) + C \sin(2\phi). \nonumber 	 
\end{eqnarray} 
Note that in the $\ell+\pi/\rho$ channels, when constructing the Optimal Observable, the internal degrees of freedom between the two neutrinos are integrated out in the corresponding coefficients, leaving only the combined four-momentum of the di-neutrino in the expressions. In the $\rho+\rho$ channel, the two neutral pions are taken to be distinguishable.

\subsection*{\texorpdfstring{$\ell+\pi$}{l+pi} channel}
We have
\begin{eqnarray}
&&A_{\pi\ell} = 2m_{\tau}^2(p_{\ell} \cdot  p_{\nu_{\tau}^{\ell}})((m_h^2 - 2m_{\tau}^2)( m_{\tau}^2-m_{\pi}^2 )(p_{\nu_{\ell}} \cdot  p_{\tau^{\ell}})\nonumber \\
&&-4m_{\tau}^4(p_{\nu_{\ell}} \cdot  p_{\pi}))\nonumber \\
&&B_{\pi\ell} = 4m_{\tau}^4(p_{\ell} \cdot  p_{\nu_{\tau}^{\ell}})((m_h^2 - 2m_{\tau}^2)(p_{\nu_{\ell}} \cdot  p_{\pi})  \nonumber \\
&& - (m_{\tau}^2 - m_{\pi}^2 )(p_{\nu_{\ell}} \cdot  p_{\tau^{\ell}}) - 2(p_{\nu_{\ell}} \cdot  p_{\tau^{\pi}})(p_{\pi} \cdot  p_{\tau^{\ell}})) \nonumber \\
&&C_{\pi\ell} = 8m_{\tau}^4
 (p_{\ell} \cdot  p_{\nu_{\tau}^{\ell}})\epsilon^{p_{\nu_{\ell}}  p_{\pi}  p_{\tau^{\ell}}  p_{\tau^{\pi}}  } 
\end{eqnarray}

\subsection*{\texorpdfstring{$\ell+\rho$}{l+rho} channel}
We have
\begin{eqnarray}
&&A_{\rho\ell} = 4(p_{\ell} \cdot  p_{\nu_{\tau}^{\ell}})(2m_{\tau}^4q_{\pi}^2(p_{\nu_{\ell}} \cdot  p_{\nu_{\tau}^{\rho}}) \nonumber \\
&&-  4m_{\tau}^4(p_{\nu_{\ell}} \cdot  q_{\pi})(p_{\nu_{\tau}^{\rho}} \cdot  q_{\pi})  + (m_h^2 - 2m_{\tau}^2)(p_{\nu_{\ell}} \cdot  p_{\tau^{\ell}})\nonumber \\
&&\times(2(p_{\nu_{\tau}^{\rho}} \cdot  q_{\pi})^2 - q_{\pi}^2(p_{\nu_{\tau}^{\rho}} \cdot  p_{\tau^{\rho}}))) \nonumber \\
&&B_{\rho\ell} = -4m_{\tau}^2(p_{\ell} \cdot  p_{\nu_{\tau}^{\ell}})(m_h^2q_{\pi}^2(p_{\nu_{\ell}} \cdot  p_{\nu_{\tau}^{\rho}}) -  2m_{\tau}^2q_{\pi}^2(p_{\nu_{\ell}} \cdot  p_{\nu_{\tau}^{\rho}})\nonumber \\
&& - 2(m_h^2 - 2m_{\tau}^2)(p_{\nu_{\ell}} \cdot  q_{\pi})   (p_{\nu_{\tau}^{\rho}} \cdot  q_{\pi})  - 2q_{\pi}^2(p_{\nu_{\ell}} \cdot  p_{\tau^{\rho}})(p_{\nu_{\tau}^{\rho}} \cdot  p_{\tau^{\ell}}) \nonumber \\
&&+ (p_{\nu_{\ell}} \cdot  p_{\tau^{\ell}})(4(p_{\nu_{\tau}^{\rho}} \cdot  q_{\pi})^2 - 2q_{\pi}^2(p_{\nu_{\tau}^{\rho}} \cdot  p_{\tau^{\rho}})) \nonumber \\
&&+ 
 4(p_{\nu_{\ell}} \cdot  p_{\tau^{\rho}})(p_{\nu_{\tau}^{\rho}} \cdot  q_{\pi})(q_{\pi} \cdot  p_{\tau^{\ell}})) \nonumber \\
&&C_{\rho\ell} = 8m_{\tau}^2(p_{\ell} \cdot  p_{\nu_{\tau}^{\ell}})
  (2(p_{\nu_{\tau}^{\rho}} \cdot  q_{\pi})\epsilon^{p_{\nu_{\ell}}  q_{\pi}  p_{\tau^{\ell}}  p_{\tau^{\rho}}  }\nonumber \\
  &&-q_{\pi}^2\epsilon^{p_{\nu_{\ell}}  p_{\nu_{\tau}^{\rho}}  p_{\tau^{\ell}}  p_{\tau^{\rho}}  })
\end{eqnarray}


\subsection*{\texorpdfstring{$\pi+\rho$}{pi+rho} channel}
We have
\begin{eqnarray}
&&A_{\pi\rho} = \frac{m_{\tau}^6}{4}(4q_{\pi}^2(p_{\nu_{\tau}^{\rho}} \cdot  p_{\pi}) - 
 8(p_{\nu_{\tau}^{\rho}} \cdot  q_{\pi})(p_{\pi} \cdot  q_{\pi}) \nonumber \\
 &&+ (\frac{m_h^2}{m_{\tau}^2} - 2)(1-\frac{m_{\pi}^2}{m_{\tau}^2} )
 (2(p_{\nu_{\tau}^{\rho}} \cdot  q_{\pi})^2 - q_{\pi}^2(p_{\nu_{\tau}^{\rho}} \cdot  p_{\tau^{\rho}})) )\nonumber \\
&&B_{\pi\rho} = -\frac{m_{\tau}^4}{2}( q_{\pi}^2((m_h^2 - 2m_{\tau}^2)(p_{\nu_{\tau}^{\rho}} \cdot  p_{\pi}) \nonumber \\
&&+ (m_{\pi}^2 - m_{\tau}^2)
       (p_{\nu_{\tau}^{\rho}} \cdot  p_{\tau^{\rho}}) - 2(p_{\nu_{\tau}^{\rho}} \cdot  p_{\tau^{\pi}})(p_{\pi} \cdot  p_{\tau^{\rho}}))    \nonumber \\
       &&  -2(m_{\pi}^2 - m_{\tau}^2)(p_{\nu_{\tau}^{\rho}} \cdot  q_{\pi})^2 + 
    (p_{\nu_{\tau}^{\rho}} \cdot  q_{\pi})(4(p_{\pi} \cdot  p_{\tau^{\rho}})(q_{\pi} \cdot  p_{\tau^{\pi}})\nonumber \\
    &&-2(m_h^2 - 2m_{\tau}^2)(p_{\pi} \cdot  q_{\pi})))\nonumber \\
&&C_{\pi\rho} = m_{\tau}^4(q_{\pi}^2\epsilon^{p_{\nu_{\tau}^{\rho}}  p_{\pi}  p_{\tau^{\pi}}  p_{\tau^{\rho}}  } \nonumber \\
&&+    2(p_{\nu_{\tau}^{\rho}} \cdot  q_{\pi})\epsilon^{p_{\pi}  q_{\pi}  p_{\tau^{\pi}}  p_{\tau^{\rho}}  })
\end{eqnarray}

\subsection*{\texorpdfstring{$\rho+\rho$}{rho+rho} channel}
We have
\begin{eqnarray}
&&A_{\rho\rho} =\frac{1}{2} (m_h^2 - 2m_{\tau}^2)(2(p_{\nu_{\tau^-}} \cdot  q_{\pi^{-}})^2 \nonumber \\
&&- q_{\pi^{-}}^2(p_{\nu_{\tau^-}} \cdot  p_{\tau^{-}}))(2(p_{\nu_{\tau^+}} \cdot  q_{\pi^{+}})^2 - q_{\pi^{+}}^2(p_{\nu_{\tau^+}} \cdot  p_{\tau^{+}}))\nonumber \\
&&- 
4m_{\tau}^4(p_{\nu_{\tau^-}} \cdot  q_{\pi^{-}})(p_{\nu_{\tau^+}} \cdot  q_{\pi^{+}})(q_{\pi^{-}} \cdot  q_{\pi^{+}})\nonumber \\
&&-m_{\tau}^4q_{\pi^{-}}^2q_{\pi^{+}}^2(p_{\nu_{\tau^-}} \cdot  p_{\nu_{\tau^+}})   + 2m_{\tau}^4q_{\pi^{+}}^2(p_{\nu_{\tau^-}} \cdot  q_{\pi^{-}})\nonumber \\
&&\times
  (p_{\nu_{\tau^+}} \cdot  q_{\pi^{-}}) + 2m_{\tau}^4q_{\pi^{-}}^2(p_{\nu_{\tau^-}} \cdot  q_{\pi^{+}})(p_{\nu_{\tau^+}} \cdot  q_{\pi^{+}}) \nonumber \\
&&B_{\rho\rho} = -\frac{m_{\tau}^2}{2}( 2m_h^2(q_{\pi^{+}}^2(p_{\nu_{\tau^-}} \cdot  q_{\pi^{-}})
     (p_{\nu_{\tau^+}} \cdot  q_{\pi^{-}}) \nonumber \\
     &&+ 
    q_{\pi^{-}}^2(p_{\nu_{\tau^-}} \cdot  q_{\pi^{+}})(p_{\nu_{\tau^+}} \cdot  q_{\pi^{+}}))    - 4m_{\tau}^2(q_{\pi^{+}}^2(p_{\nu_{\tau^-}} \cdot  q_{\pi^{-}})\nonumber \\
    &&\times(p_{\nu_{\tau^+}} \cdot  q_{\pi^{-}})  + q_{\pi^{-}}^2(p_{\nu_{\tau^-}} \cdot  q_{\pi^{+}})(p_{\nu_{\tau^+}} \cdot  q_{\pi^{+}})) \nonumber \\
    &&+  8(p_{\nu_{\tau^-}} \cdot  q_{\pi^{-}})^2(p_{\nu_{\tau^+}} \cdot  q_{\pi^{+}})^2     -(m_h^2-2m_{\tau}^2)q_{\pi^{-}}^2q_{\pi^{+}}^2\nonumber \\
    &&\times(p_{\nu_{\tau^-}} \cdot  p_{\nu_{\tau^+}})
     + 2q_{\pi^{-}}^2q_{\pi^{+}}^2((p_{\nu_{\tau^-}} \cdot  p_{\tau^{+}}) 
     (p_{\nu_{\tau^+}} \cdot  p_{\tau^{-}}) \nonumber \\
     &&+ (p_{\nu_{\tau^-}} \cdot  p_{\tau^{-}})(p_{\nu_{\tau^+}} \cdot  p_{\tau^{+}})) -    4q_{\pi^{-}}^2(p_{\nu_{\tau^-}} \cdot  p_{\tau^{+}})\nonumber \\
     &&\times (p_{\nu_{\tau^+}} \cdot  q_{\pi^{+}})(q_{\pi^{+}} \cdot  p_{\tau^{-}}) -  4q_{\pi^{+}}^2(p_{\nu_{\tau^-}} \cdot  q_{\pi^{-}})\nonumber \\
     &&\times(p_{\nu_{\tau^+}} \cdot  p_{\tau^{-}})(q_{\pi^{-}} \cdot  p_{\tau^{+}}) )     + 
    8(p_{\nu_{\tau^-}} \cdot  q_{\pi^{-}})(p_{\nu_{\tau^+}} \cdot  q_{\pi^{+}})\nonumber \\
    &&\times(m_{\tau}^2(q_{\pi^{-}} \cdot  q_{\pi^{+}}) + (q_{\pi^{-}} \cdot  p_{\tau^{+}})(q_{\pi^{+}} \cdot  p_{\tau^{-}}))     \nonumber \\
    && - 4(q_{\pi^{+}}^2(p_{\nu_{\tau^-}} \cdot  q_{\pi^{-}})^2(p_{\nu_{\tau^+}} \cdot  p_{\tau^{+}}) \nonumber \\
    &&+ q_{\pi^{-}}^2(p_{\nu_{\tau^+}} \cdot  q_{\pi^{+}})^2(p_{\nu_{\tau^-}} \cdot  p_{\tau^{-}})
     )    \nonumber \\
     &&- 4m_h^2(p_{\nu_{\tau^-}} \cdot  q_{\pi^{-}})(p_{\nu_{\tau^+}} \cdot  q_{\pi^{+}})(q_{\pi^{-}} \cdot  q_{\pi^{+}}) \nonumber \\
&&C_{\rho\rho} = m_{\tau}^2(\epsilon^{p_{\nu_{\tau^-}}  p_{\nu_{\tau^+}}  p_{\tau^{-}}  p_{\tau^{+}}  }q_{\pi^{-}}^2q_{\pi^{+}}^2 \nonumber \\
&&+ 
   2\epsilon^{p_{\nu_{\tau^+}}  q_{\pi^{-}}  p_{\tau^{-}}  p_{\tau^{+}}  }q_{\pi^{+}}^2
    (p_{\nu_{\tau^-}} \cdot  q_{\pi^{-}}) \nonumber \\
&&
    - 2\epsilon^{p_{\nu_{\tau^-}}  q_{\pi^{+}}  p_{\tau^{-}}  p_{\tau^{+}}  }
    q_{\pi^{-}}^2(p_{\nu_{\tau^+}} \cdot  q_{\pi^{+}})\nonumber \\
    && + 4\epsilon^{q_{\pi^{-}}  q_{\pi^{+}}  p_{\tau^{-}}  p_{\tau^{+}}  }
    (p_{\nu_{\tau^-}} \cdot  q_{\pi^{-}})(p_{\nu_{\tau^+}} \cdot  q_{\pi^{+}}))
\end{eqnarray}

\section{The definitions of \texorpdfstring{$\Delta\phi_{\text{IP}}$}{phiIP} and \texorpdfstring{$\Delta\phi_{\text{CP}}$}{phiCP} in \texorpdfstring{$\rho+\rho$}{rho+rho} channel}
\label{appPhi}
In the lab frame, the reconstructed four-momentums for $\pi^+$, $\pi^-$, $\pi^0_+$, $\pi^0_-$ are $p_{\pi^+}$, $p_{\pi^-}$, $p_{\pi^0_+}$, $p_{\pi^0_-}$, and the impact parameter vectors for $\tau^+$ and $\tau^-$ are $\vec{n}^+$ and $\vec{n}^-$ from which we can construct the four-vector for each impact parameter: $n^+=(0,\vec{n}^+)$ and $n^-=(0,\vec{n}^-)$. 

Using four four-vectors ($p_{m1}$, $p_{d1}$, $p_{m2}$, $p_{d2}$), we first define the zero momentum frame (ZMF) of $p_{m1}$ and $p_{m2}$. In this frame, the four-vectors, after a Lorentz transformation, become ($\hat{p}_{m1}$, $\hat{p}_{d1}$, $\hat{p}_{m2}$, $\hat{p}_{d2}$). Then we consider the angle defined by
\begin{eqnarray}
\label{equ:phi}
\Phi = \arccos(\hat{p}_{d1}^{\perp}\cdot\hat{p}_{d2}^{\perp})\times\text{sgn}(\hat{p}_{m1}\cdot(\hat{p}_{d1}^{\perp}\times\hat{p}_{d2}^{\perp}))
\end{eqnarray}
where the superscript $\perp$ denotes the perpendicular components of the momentum transverse to $\hat{p}_{m1}$ ($\hat{p}_{m2}$).

$\Delta\phi_{\text{IP}}$ is built by Eq.~\ref{equ:phi} with the input four-vectors as ($p_{\pi^+}+p_{\pi^0_+}$, $n^+$, $p_{\pi^-}+p_{\pi^0_-}$, $n^-$), while the input four-vectors for $\Delta\phi_{\text{CP}}$ are ($p_{\pi^+}$, $p_{\pi^0_+}$, $p_{\pi^-}$, $p_{\pi^0_-}$). Further, for $\Delta\phi_{\text{CP}}$ of $\rho+\rho$ channel, there is a subtlety associated with the vertex structure of $\rho\pi\pi$ which is proportional to ($p_{\pi^\pm}-p_{\pi^0}$), hence we need to redefine the angle according to $Y=(E_{\pi^+}-E_{\pi^0_+})(E_{\pi^-}-E_{\pi^0_-})$ as:
\begin{eqnarray}
\Delta\phi_{CP} =\begin{cases}
\Phi - \pi, \quad \text{if } Y<0\text{ and } 0\leq\Phi\leq\pi, \\
\Phi + \pi, \quad \text{if } Y<0\text{ and } -\pi\leq\Phi<0,\\
\Phi,\quad\qquad \text{otherwise}.
\end{cases}
\end{eqnarray}

\begin{acknowledgement}

The authors would like to thank the support from Tsinghua University, Center of High Energy Physics of Tsinghua University, Collaborative Innovation Center of Quantum Matter of China, Center for High Energy Physics of Peking University, the National Thousand Young Talents program and the NSFC of China.
\end{acknowledgement}
%
\bibliographystyle{bibsty}
\bibliography{references}
%

\end{document}